\documentclass[sigconf, nonacm]{acmart} 

\AtBeginDocument{%
  }

\usepackage{subcaption}
\usepackage{multirow}

\usepackage{algorithm}
\usepackage{algorithmic}

\newsavebox{\takeawaybin}
\newenvironment{takeawaybox}
  {\par\smallskip\noindent
   \begin{lrbox}{\takeawaybin}%
   \begin{minipage}{\dimexpr\linewidth-13pt\relax}%
   \setlength{\parindent}{0pt}}
  {\end{minipage}\end{lrbox}%
   \fboxsep=4pt\fboxrule=0.5pt%
   \fcolorbox{black}{gray!10}{\hspace{2pt}\usebox{\takeawaybin}\hspace{2pt}}%
   \par\smallskip}

\begin{document}


\title{Position Matters: Feature Inversion Attacks in ViT Split Inference with Token Reduction and Shuffling} 



\author{Stefano Leggio}
\email{stefano.leggio@santannapisa.it}
\affiliation{%
  \institution{Dept. of Excellence in Robotics \& AI, Scuola Superiore Sant'Anna}
  \city{Pisa}
  \country{Italy}
}

\author{Giulio Rossolini}
\email{giulio.rossolini@santannapisa.it}
\affiliation{%
  \institution{Dept. of Excellence in Robotics \& AI, Scuola Superiore Sant'Anna}
  \city{Pisa}
  \country{Italy}
}

\author{Alessandro Biondi}
\email{alessandro.biondi@santannapisa.it}
\affiliation{%
  \institution{Dept. of Excellence in Robotics \& AI, Scuola Superiore Sant'Anna}
  \city{Pisa}
  \country{Italy}
}

\begin{abstract}
Vision Transformers (ViTs) are increasingly used in split-inference systems, where edge devices transmit intermediate token representations to a remote cloud. In this setting, token reduction lowers computation and communication costs, while token shuffling disrupts the spatial organization of the transmitted tokens, potentially limiting information leakage. However, their privacy benefits remain unclear against feature inversion attacks, which attempt to reconstruct the input from the transmitted embeddings.
In this work, we show that, despite disrupting the spatial structure required by conventional reconstruction attacks, transmitted token embeddings retain substantial positional information. Based on this observation, we introduce the \emph{Spatially Aligned Reconstruction Attack} (SARA), a unified pipeline that predicts token positions, restores their spatial layout, reconstructs missing embeddings using a feature-space masked autoencoder, and recovers the input image. Our results demonstrate that token shuffling provides only apparent privacy, as SARA largely reconstructs the original token organization. Token reduction offers stronger protection, but significant leakage persists when the retained tokens preserve sufficient semantic and positional information.
Finally, we introduce a lightweight edge-side defense that removes positional embeddings and progressively adapts the edge-side transformer blocks through knowledge distillation. It substantially reduces attack performance against SARA, while preserving downstream task accuracy and requiring no changes to the cloud-side model.
\end{abstract}







\maketitle

\section{Introduction}
\label{sec:introduction}
Vision Transformers (ViTs)~\cite{Dosovitskiy21} have become a reference architecture for computer vision, achieving strong performance across a wide range of tasks~\cite{liu2022surveyvit, clip_radford2021learning, simeoni2025dinov3}. ViTs process an image by splitting it into a sequence of patch tokens, each encoding the visual content of a local image region together with information about its spatial position. This sequence-based representation makes intermediate embeddings naturally well suited to token-level manipulations, often without modifying the model parameters or retraining the downstream network. For example, \emph{token reduction} methods~\cite{rao2021dynamicvit, bolya2023tome, kim2024tofu, papa2024efficientvit} drop or merge redundant and less informative tokens, thereby reducing the computational cost of subsequent Transformer blocks while preserving a favorable trade-off with task performance. In contrast, \emph{token shuffling} methods~\cite{yao2022patchshuffling, xu2024permutation, li2026shuffling} permute the token sequence, disrupting the explicit correspondence between individual tokens and the spatial layout of the input image. Although these operations are primarily introduced for efficiency or representation manipulation, they may also provide an apparent privacy benefit by limiting the amount of information exposed in intermediate features and/or concealing the correspondence between transmitted tokens and their original image locations.

\begin{figure*}[t]
    \centering
    \includegraphics[width=1\textwidth]{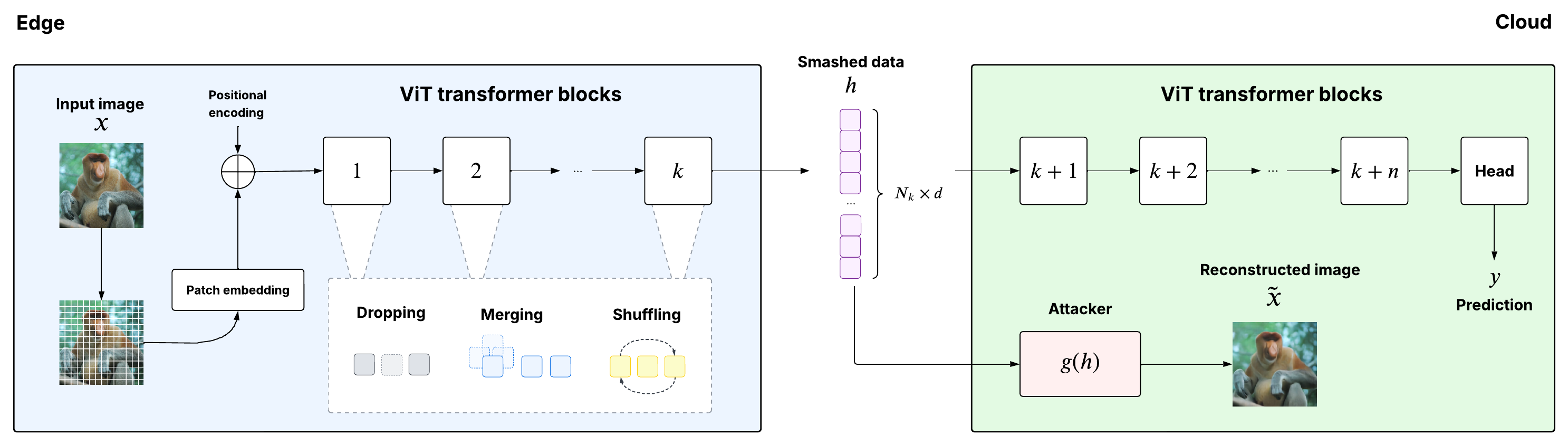}
    \caption{\small{Illustration of the threat model. The edge device transmits the smashed data $h$ to the cloud, where an honest-but-curious attacker attempts to output a reconstruction of the original input image $\tilde{x}$.}}
    \label{fig:threat_model}
\end{figure*}

The combination of efficiency and potential privacy benefits makes these operations particularly appealing in \emph{split inference} scenarios~\cite{kang2017neurosurgeon, iot_ROSSOLINI2025101795, he2021attacking}, where a neural network is partitioned between an edge device holding the private input and a remote cloud. In privacy-sensitive applications, the edge device processes the input through the initial portion of the model and transmits the resulting intermediate representation to the cloud, which completes the inference. 
Token manipulations can therefore be naturally applied on the edge side before transmission, reducing client-side computation and communication costs while potentially obscuring the spatial correspondence between the transmitted tokens and the original image regions. In this setting, it is reasonable to consider an \emph{honest-but-curious} threat model \cite{luo2024llrq, he2019modelinversion, he2021attacking}, in which the cloud correctly performs the assigned inference computation but may also attempt to reconstruct the private input from the received features, as illustrated in Figure~\ref{fig:threat_model}.

However, the privacy provided by these token manipulations remains insufficiently investigated, despite being of crucial importance in split-inference scenarios. Conventional feature inversion attacks typically assume that intermediate representations preserve a fixed spatial organization~\cite{he2019modelinversion, he2021attacking, yang2022measuring, lei2025drag, Erak25, khan2026splitattacks}. Consequently, their failure after token reduction or token shuffling may simply result from the violation of this assumption, rather than from a genuine reduction in information leakage. Indeed, the transmitted token embeddings may still retain sufficient semantic and positional information to recover their original spatial organization and reconstruct the private input. This motivates the need for reconstruction attacks specifically designed to operate on manipulated token representations.



\textbf{Paper contributions.} In this work, we provide a systematic privacy-oriented analysis of token reduction and token shuffling in ViT-based split inference. We first show that intermediate token embeddings retain substantial positional information, even after passing through multiple ViT blocks, which enables a position-aware attacker to infer their original spatial location. Building on this observation, we introduce the \emph{Spatially Aligned Reconstruction Attack} (SARA), a unified reconstruction pipeline that predicts the original positions of the transmitted tokens, restores their spatial arrangement, reconstructs missing token embeddings through a feature-space masked autoencoder, and finally recovers the input image. This pipeline enables a consistent evaluation of information leakage under both token reduction and token shuffling.

Experimental results show that token shuffling alone provides only a false sense of privacy, as its apparent benefits can be greatly disrupted by SARA. Token reduction offers stronger protection by discarding part of the representation, but introduces a privacy--utility trade-off due to the resulting degradation in task accuracy. Nevertheless, its privacy benefits remain limited when the retained tokens preserve sufficient semantic and positional information for SARA to infer the missing content.

Motivated by these findings, we finally introduce a lightweight edge-side defense, specifically tailored to split inference settings, where adapting the cloud-side model should be avoided to accommodate multiple edge devices. The proposed mechanism provides a simple yet effective baseline that attenuates positional cues before transmission, and substantially reduces reconstruction leakage against ad-hoc attacks, while maintaining a favorable trade-off with task performance.


\textbf{Paper Structure.} The remainder of the paper is organized as follows. Section~\ref{sec:preliminaries} introduces the threat model for ViT-based split inference and reviews the token manipulation strategies considered in this work. Section~\ref{sec:attack_pipeline} presents the proposed SARA attack pipeline and provides a detailed description of its components. Section~\ref{sec:positionless_adaptation} introduces a simple yet effective edge-side defense. Section~\ref{sec:experiments} first evaluates the effectiveness of SARA against the considered token manipulation strategies and then assesses the benefits of the proposed defense. Finally, Sections~\ref{sec:related_work} and~\ref{sec:conclusion} discuss the connections with the related literature and present the conclusions and limitations, respectively.
The source code and experimental implementation are publicly available at the following repository:
\url{https://github.com/stenaflow/position-matters}.

\section{Preliminaries and Threat Model}
\label{sec:preliminaries}
This section formalizes the split-inference setting under consideration, the token-manipulation operations applied at the edge, and the capabilities and objective of the adversary.

\subsection{ViT tokenization and token operations} 
\label{sec:token_operations} 
Let \(x\in[0,1]^{C\times H\times W}\) be an input image. A ViT \(f\) partitions \(x\) into \(
N_0=\frac{HW}{p^2}
\) non-overlapping image patches of size \(p\times p\).
Each patch is then linearly projected into a \(d\)-dimensional embedding and combined with a learnable positional embedding (PE). The patch input sequence to the first transformer block is therefore
\begin{equation}
{h}^{(0)}
=
\left[
{h}^{(0), 1};
\ldots;
{h}^{(0), N_0}
\right]
\in
\mathbb{R}^{N_0\times d},
\label{eq:vit_input_sequence}
\end{equation} 
where \({h}^{(0),i}\in\mathbb{R}^{d}\) denotes the embedding associated with
the \(i\)-th image patch, including its positional embedding. Note that, from the sequence ${h}^{(0)}$, we intentionally omit the classification token to denote only the patch-token representations.\footnote{The classification token is
still processed by all ViT layers and is included in the legitimate inference
pipeline. We omit it from the reconstruction formulation because it does not
have a direct correspondence with a spatial image patch and is not explicitly
used by the considered reconstruction attack and pipelines.}

From a layer-level perspective, let the ViT consist of \(L\) blocks \(\{{B}_1,\ldots,{B}_L\}\). In the original ViT architecture, without additional token-manipulation operations, each block preserves the sequence length such that the patch-token representation produced after block \(B_\ell\) is
\(
{h}^{(\ell)}
\in
\mathbb{R}^{N_\ell\times d},
\)
with \(N_\ell=N_0\) for every \(\ell\in\{1,\ldots,L\}\).

\emph{Token-reduction} methods modify ViT inference by progressively decreasing the
number of tokens processed by subsequent transformer blocks. In particular, operations such as token dropping~\cite{rao2021dynamicvit} and token merging~\cite{bolya2023tome} remove or aggregate selected tokens, respectively.
Consequently, the sequence length becomes \(N_i\leq N_j\) for \(i>j\).
For simplicity, we first consider a fixed token reduction policy, where \(r\) denotes the number of patch tokens removed from the sequence, either by dropping or merging, after each transformer block output.
Assuming that the same reduction amount \(r\) is applied after each of the first \(\ell\) transformer blocks, the number of patch tokens available in \(h^{(\ell)}\) is
\(
N_{\ell}=\max\{0,\,N_0-\ell r\},
\)
where \(N_0\) denotes the initial number of patch tokens. Once \(N_{\ell}=0\), only the class token remains, and no further token reduction is applied in the subsequent blocks. Importantly, increasing $r$ improves computational and communication efficiency by reducing the number of processed and transmitted tokens. However, stronger reduction may improve privacy at the cost of downstream accuracy. To capture this trade-off, in Section~\ref{sec:experiments} we introduce a unified metric by jointly accounting for task utility and privacy leakage.

In contrast, \emph{token shuffling} does not change the number of transmitted tokens.
Instead, it permutes their sequence order at the output of a selected block \(\ell\). Thus, when shuffling is applied without token reduction, it holds \(N_\ell=N_0\), while the representation transmitted after block \(\ell\) is an ordered permutation of the original token sequence.

\subsection{ViT Split Inference}
We consider a split point \(k\), with \(1\leq k<L\), such that the edge executes the first \(k\) transformer blocks and produces the intermediate representation
\(
{h}^{(k)} = f_{\mathrm{e}}^{(k)}(x),
\)
while the cloud executes the remaining \(L-k\) blocks and returns the predicted logits
\(
f_{\mathrm{c}}^{(k)}
\left(
{h}^{(k)}
\right).
\)
Consequently, when token reduction or shuffling is applied on the edge-side, the corresponding operations are included in \(f_{\mathrm{e}}^{(k)}\), while the learned ViT parameters remain unchanged. The cloud-side model therefore receives the manipulated token sequence and continues the inference procedure from the selected split point.
As discussed in Section~\ref{sec:introduction}, these techniques are particularly relevant to split inference because they can improve computational and communication efficiency and, in the case of token shuffling, conceal the original spatial arrangement of the transmitted tokens. 

\subsection{Threat Model}
\label{sec:threat_model}
We consider an \emph{honest-but-curious} cloud that correctly executes the cloud-side portion of the ViT but attempts to infer private information about the edge's input. For an input image \(x\), the adversary observes the intermediate representation
\(
{h}^{(k)}
=
f_{\mathrm{e}}^{(k)}(x), 
\)
with $N_k$ token embeddings, transmitted at split point \(k\). Depending on the edge-side configuration, \({h}^{(k)}\) may contain a reduced token sequence and/or a shuffled token sequence. 
Note that, in this threat model, the adversary cannot directly observe private edge-side information associated with a specific input, such as the original image, the spatial provenance of the retained or merged tokens, or the permutation used to shuffle the transmitted sequence. 
The adversary is passive and does not alter the inference protocol, the transmitted representation, or the legitimate prediction. Its objective is to reconstruct the private input from the observed smashed data. 

Following these assumptions, the attacker trains a reconstruction model
\(
g_{\theta}:
\mathbb{R}^{N_k\times d}
\rightarrow
[0,1]^{C\times H\times W}
\)
using an auxiliary dataset
\(
\mathcal{D}_{\mathrm{aux}}
=
\left\{
\left(x_i,{h}^{(k)}_i\right)
\right\}_{i=1}^{M},
\)
where parameters are learned by solving
\begin{equation}
\theta_g^{*}
=
\arg\min_{\theta_g}
\frac{1}{M}
\sum_{i=1}^{M}
\mathcal{L}_{\mathrm{rec}}
\left(
g_{\theta}\!\left({h}^{(k)}_i\right),
x_i
\right),
\label{eq:reconstruction_objective}
\end{equation}
where \(\mathcal{L}_{\mathrm{rec}}\) denotes the reconstruction loss, e.g., a pixel-level mean square error (MSE). 
After training $g$, given a smashed representation of an unseen private input, the attacker produces
\(
\tilde{x}
=
g\!\left({h}^{(k)}\right).
\)
We assess privacy leakage through the similarity between \(x\) and \(\tilde{x}\), measured using pixel-level and perceptual reconstruction metrics (see Section \ref{sec:experimental_settings}). Higher reconstruction quality indicates that more information about the private input can be recovered from the transmitted token representation. 
\section{Attack Pipeline}
\label{sec:attack_pipeline}
This section describes the Spatially Aligned Reconstruction Attack (SARA). We first motivate the idea by highlighting the importance of recovering positional information. We then present the steps as a unified approach for reconstructing images from features that have undergone token reduction or shuffling on the edge side.

\subsection{Motivation and Intuition}
Standard reconstruction attacks on intermediate representations typically train a convolutional decoder to recover the input image directly from the transmitted data $h^{(k)}$~\cite{he2019modelinversion, he2021attacking, yang2022measuring, lei2025drag, Erak25, khan2026splitattacks}. These approaches generally assume that the intermediate representation preserves a fixed spatial organization that can be mapped back to the original image via classic operations. However, token reduction and shuffling violate this assumption by altering the number or ordering of the transmitted tokens, potentially creating a misleading impression of improved privacy when evaluated against convolutional-based reconstructions.

\begin{figure}[t]
    \centering
    \begin{subfigure}{0.98\columnwidth}
        \centering
        \small
        \resizebox{\columnwidth}{!}{%
        \begin{tabular}{c cccccccccccc c}
            \toprule
            \textbf{Split point} &
            \textbf{1} & \textbf{2} & \textbf{3} & \textbf{4} &
            \textbf{5} & \textbf{6} & \textbf{7} & \textbf{8} &
            \textbf{9} & \textbf{10} & \textbf{11} & \textbf{Avg.} \\
            \midrule
            \textbf{Aligned}
            & 0.92 & 0.89 & 0.84 & 0.79
            & 0.74 & 0.69 & 0.65 & 0.61
            & 0.56 & 0.53 & 0.49 & \textbf{0.70} \\
        
            \textbf{Shuffled}
            & 0.18 & 0.21 & 0.22 & 0.24
            & 0.25 & 0.26 & 0.27 & 0.27
            & 0.27 & 0.27 & 0.27 & \textbf{0.25} \\
            \bottomrule
        \end{tabular}
        }
    \end{subfigure}

    \vspace{1em}

    \includegraphics[width=0.8\columnwidth]{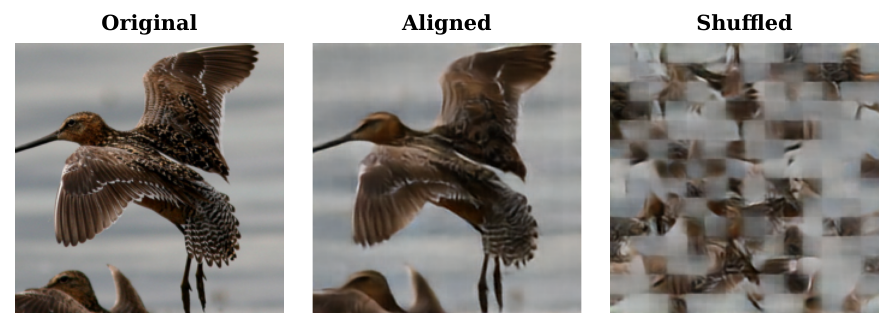}

    \vspace{1em}
    
    \begin{subfigure}{\columnwidth}
        \centering
        \includegraphics[width=0.80\columnwidth]{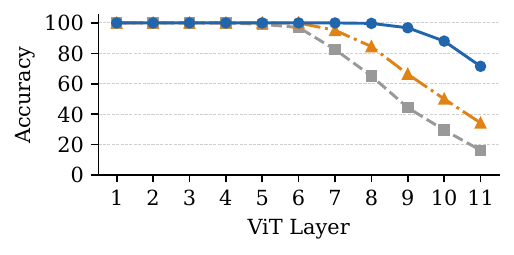}
    \end{subfigure}

    \begin{subfigure}{\columnwidth}
        \centering
        \includegraphics[width=0.75\columnwidth]{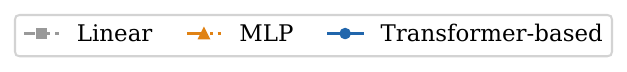}
    \end{subfigure}
    \vspace{-2em}
    \caption{\small{(Top) SSIM achieved by a conventional convolutional decoder on ImageNet when the transmitted ViT-B tokens retain their original spatial order (\emph{Aligned}) or are randomly permuted (\emph{Shuffled}); (center) illustration of the original input and reconstructions at split point 4 with a convolutional decoder; (bottom) Top-1 token-position accuracy across ViT layers for different predictors.}}
    \label{fig:motivation}
\end{figure}

To illustrate this limitation, we evaluate a standard reconstruction decoder under two configurations of ViT-B/16 (setup details in Section \ref{sec:experimental_settings}): an \emph{aligned} setting, in which tokens preserve their original spatial order, and a \emph{shuffled} setting, in which their order is randomly permuted. As reported in the table in Figure \ref{fig:motivation} (top), shuffling reduces the average Structural Similarity
Index Measure (SSIM)~\cite{wang2004ssim} from $0.700$ to $0.246$. As expected, a consistent drop across all split points confirms that conventional reconstruction decoders strongly depend on the original token arrangement. However, shuffling only changes the order of the transmitted tokens; it does not directly remove the semantic or positional information encoded in their representations.
This observation naturally raises the question of \emph{whether shuffling operations can be inverted by inferring the original spatial position of each token}. In principle, recovering the corresponding patch index would allow the tokens to be rearranged before reconstruction, thereby restoring the spatial structure and recovering the same reconstruction quality as in the aligned setting. 

To investigate this possibility, we train different token-position predictors to classify each intermediate token according to its original position in the input patch grid. As shown in Figure~\ref{fig:motivation} (bottom), positional information remains highly accessible throughout the ViT. Even a linear predictor achieves nearly perfect accuracy in the early layers, although its performance decreases substantially for deeper representations. In contrast, the Transformer-based predictor maintains high accuracy through the layers.

\subsection{Spatially Aligned Reconstruction Attack}
\label{sec:sara}

\begin{figure*}[t]
    \centering
    \includegraphics[width=1.0\textwidth]{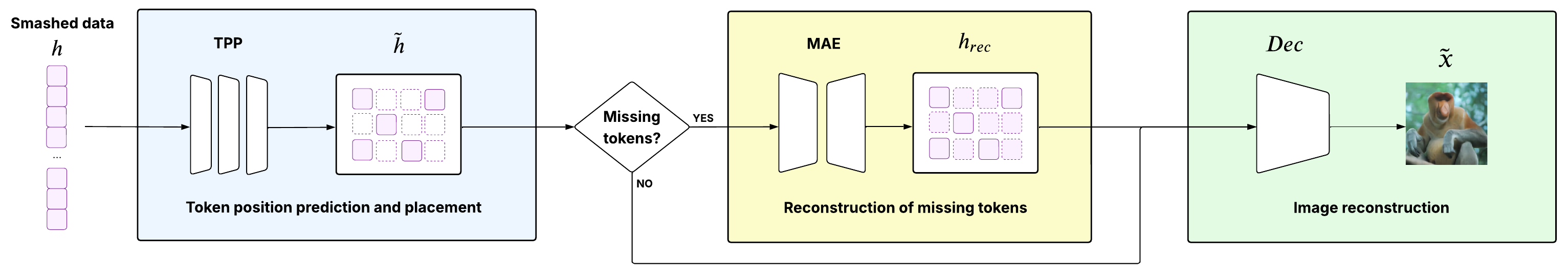}
    \caption{\small{Overview of the Spatially Aligned Reconstruction Attack (SARA). The pipeline consists of three stages: (1) token position prediction and spatial alignment, (2) missing-token reconstruction via a Masked Autoencoder, and (3) image reconstruction with a convolutional decoder.}}
    \label{fig:attack_pipeline}
\end{figure*}

We propose the \emph{Spatially Aligned Reconstruction Attack} (SARA), a three-stage pipeline designed to improve the reconstruction of smashed representations $h^{(k)}$ affected by token reduction or shuffling. The overall pipeline is illustrated in Figure~\ref{fig:attack_pipeline} and consists of three stages, which are described more technically as follows: (i)~a \emph{Token Position Predictor}, which estimates the original spatial position of each transmitted token and constructs a spatially aligned, full-length masked representation; (ii)~a \emph{Masked Autoencoder}, which imputes the representations associated with missing token positions; and (iii)~a \emph{Decoder}, which maps the restored smashed representation back to the input image through a convolutional reconstruction network.

\subsubsection{Token Position Prediction and Placement}
\label{sec:token_position_predictor}
Given an unordered set of smashed tokens, the Token Position Predictor (TPP) estimates the original image-patch position of each token and rearranges the tokens accordingly to recover their spatial organization. 

For simplicity, we omit the split-point superscript and denote the smashed representation at a split point $k$ as
\(
    h = \{h^1,\ldots,h^{N_k}\}, 
\)
\(
    h^i \in \mathbb{R}^{d},
\)
where $N_k$ is the number of tokens transmitted after token manipulation, $N_0$ is the number of image patches in the original token sequence, and $d$ is the token dimensionality.
The TPP is implemented as a position classifier $C_{\mathrm{pos}}$ that assigns each transmitted token a probability distribution over the $N_0$ possible patch positions:
\(
    P = C_{\mathrm{pos}}(h)
    \in [0,1]^{N_k \times N_0},
    \qquad
    \sum_{j=1}^{N_0} P_{i,j}=1,
\)
where $P_{i,j}$ denotes the predicted probability that token $h^i$ originated from the $j$-th image patch. For each token, we define its predicted position and the corresponding confidence as
\begin{equation}
    \hat{p}_i
    =
    \arg\max_{j \in \{1,\ldots,N_0\}} P_{i,j},
    \qquad
    c_i
    =
    \max_{j \in \{1,\ldots,N_0\}} P_{i,j}.
\end{equation}

The transmitted tokens are then placed into a full-length representation
$\tilde{h}\in\mathbb{R}^{N_0\times d}$. For each position $j$, we first identify the set of sufficiently confident tokens assigned to that position:
\begin{equation}
    \mathcal{I}_j
    =
    \left\{
        i \,\middle|\,
        \hat{p}_i=j
        \ \land\
        c_i \geq \tau
    \right\},
\label{eq:selection_thr}
\end{equation}
where $\tau$ is a confidence threshold. If multiple tokens are assigned to the same position, we retain the one with the highest confidence. The spatially aligned representation is therefore constructed as
\begin{equation}
    \tilde{h}_j =
    \begin{cases}
        h_{i_j^{*}},
        &
        \text{if } \mathcal{I}_j \neq \emptyset,
        \\[2mm]
        h_{\mathrm{void}},
        &
        \text{otherwise},
    \end{cases}
    \qquad 
    i_j^{*}
    =
    \arg\max_{i\in\mathcal{I}_j} c_i,
\end{equation}
where $h_{\mathrm{void}}=\mathbf{0}\in\mathbb{R}^{d}$ denotes the void value.
This operation restores the spatial indexing required by the subsequent reconstruction stages. Positions associated with removed tokens, low-confidence predictions, or unresolved assignments remain marked by the void value and are subsequently processed by the masked autoencoder.

For training, the attacker uses an auxiliary image dataset to extract the corresponding smashed representations and simulate random shuffling across their token order. The TPP is then optimized using a cross-entropy loss, where the ground-truth label of each token corresponds to its original patch index.

\subsubsection{Reconstruction of Missing Tokens}
\label{sec:masked_autoencoder}
The spatially aligned representation produced by the previous step, \(\tilde{h}\in\mathbb{R}^{N_0\times d}\), may still contain missing token embeddings, represented by zero vectors at their corresponding positions. Whenever such missing positions remain, we adopt the principle of masked auto-encoders (MAEs)~\cite{he2022mae} to recover the unavailable representations.

In particular, a feature-space masked autoencoder, denoted by $\mathcal{M}$, maps this incomplete representation to a complete reconstructed representation:
\(
    h_{\mathrm{rec}}
    =
    \mathcal{M}(\tilde{h})
    \in
    \mathbb{R}^{N_0\times d}.
\)

To do this, the model is trained to recover the complete smashed representation from a partially observed version of it; a subset of token embeddings in a full smashed representation, denoted for simplicity as $h_{\mathrm{full}} \in\mathbb{R}^{N_0\times d}$, collected by the attacker from the auxiliary dataset, is randomly selected and replaced with zero vectors, producing a masked representation $h_{\mathrm{mask}} \in\mathbb{R}^{N_0\times d} $ that emulates the incomplete representation ($\tilde{h}$) encountered at inference time. The model is then optimized using the MSE between the reconstructed and complete representations:
\begin{equation}
    \mathcal{L}_{\mathrm{MAE}}
    =
    \frac{1}{N_0 ~ d}
    \left\|
        \mathcal{M}(h_{\mathrm{mask}})
        -
        h_{\mathrm{full}}
    \right\|_F^2,
    \label{eq:mae_loss}
\end{equation}
where $\|\cdot\|_F$ denotes the Frobenius norm.
By learning the dependencies among the visible token embeddings, the model infers the representations associated with missing spatial positions.  When $\tilde{h}$ contains no missing positions, the masked-autoencoder stage is bypassed, and we directly set $h_{\mathrm{rec}}=\tilde{h}$.

\subsubsection{Image reconstruction}
\label{sec:decoder}
The completed smashed representation $h_{rec}$ is finally passed to an image reconstruction decoder, denoted by $\textit{Dec}$, which reconstructs an approximation of the original input image \(
\tilde{x}
=
\textit{Dec}(h_{rec}).
\)
In particular, the reconstructed patch embeddings in \(h_{\mathrm{rec}}\) are first linearly projected and reshaped into a spatial grid. A convolutional decoder composed of upsampling blocks then progressively restores the original image resolution. As commonly done in reconstruction attacks~\cite{he2019modelinversion, he2021attacking, yang2022measuring, lei2025drag, Erak25, khan2026splitattacks}, the decoder is trained on an auxiliary image dataset available to the attacker and optimized by minimizing the mean squared error between the reconstructed image and the corresponding ground-truth input.
Note that we train the decoder independently of the MAE stage, allowing us to better isolate the benefits of the preceding stages and provide a fairer comparison with conventional reconstruction approaches based solely on a decoder.

\section{Proposed Defense}
\label{sec:positionless_adaptation}
To mitigate the high effectiveness of SARA, as demonstrated by the experimental results in Section~\ref{sec:experiments}, we introduce a simple yet effective edge-side defense that limits the attacker’s ability to infer the positional information retained in the smashed representations.

The proposed defense pursues two objectives. First, it aims to attenuate the positional information encoded in the smashed representations, thereby reducing the effectiveness of the TPP stage in SARA. Second, it aims to preserve the task-level behavior of the original model without modifying the cloud-side ViT part. 
To achieve these objectives, we adopt a knowledge-distillation approach in which a student version of the original ViT is constructed by removing the positional embeddings from the edge-side model. Its transformer blocks are then progressively adapted using the original pretrained ViT as the teacher. The overall procedure is summarized in Algorithm~\ref{alg:positionless_adaptation} and described in detail below.

\begin{algorithm}[t]
\caption{Progressive Positionless Edge Finetuning}
\label{alg:positionless_adaptation}
\begin{algorithmic}[1]

\REQUIRE
Pretrained edge $f_{\mathrm{e}}^{(k)}$,
cloud model $f_{\mathrm{c}}^{(k)}$,
unlabeled adaptation dataset $\mathcal{D}_{\mathrm{A}}$,
split point $k$,
number of epochs $N_{\mathrm{epoch}}$

\STATE Initialize the student edge
$\bar{f}_{\mathrm{e}}^{(k)}
\gets
f_{\mathrm{e}}^{(k)}$
\STATE Remove its positional embeddings

\FOR{$j=1$ to $k$}
    \FOR{$e=1$ to $N_{\mathrm{epoch}}$}
        \FORALL{mini-batches $\mathcal{B}\subset\mathcal{D}_{\mathrm{A}}$}
            \STATE
            $\mathcal{L}_{\mathrm{KD}}
            \gets
            D_{\mathrm{KL}}
            \left(
               {f}_{\mathrm{c}}^{(k)}( \bar{f}_{\mathrm{e}}^{(k)}(\mathcal{B}))
                \,\middle\|\,
                {f}_{\mathrm{c}}^{(k)}( {f}_{\mathrm{e}}^{(k)}(\mathcal{B}))
            \right)$

            \STATE Update the $j$-th block of $\bar{f}_{\mathrm{e}}^{(k)}$ to minimize
            $\mathcal{L}_{\mathrm{KD}}$

        \ENDFOR
    \ENDFOR
\ENDFOR

\RETURN finetuned student edge $\bar{f}_{\mathrm{e}}^{(k)}$

\end{algorithmic}
\end{algorithm}

We use the original pretrained model as a frozen teacher, composed of the edge-side component \(f_{\mathrm{e}}^{(k)}\) and the cloud-side component \(f_{\mathrm{c}}^{(k)}\). The student differs from the teacher only in its edge-side component, denoted by \(\bar{f}_{\mathrm{e}}^{(k)}\). Before fine-tuning, \(\bar{f}_{\mathrm{e}}^{(k)}\) is initialized with the same pretrained parameters as \(f_{\mathrm{e}}^{(k)}\), but the positional embeddings stage is removed, as shown in the following. Omitting the class token for simplicity, the initial teacher and student representations are therefore given by
\[
    \mathrm{teacher}: h^{(0)}
    \leftarrow
    E_{\mathrm{patch}}(x)
    +
    E_{\mathrm{pos}},
    \qquad
    \mathrm{student}: h^{(0)}
    \leftarrow
    E_{\mathrm{patch}}(x),
\]
where \(E_{\mathrm{patch}}\) denotes the patch-embedding function and \(E_{\mathrm{pos}}\) denotes the positional embeddings.

Although directly removing the positional embeddings eliminates the primary source of positional information, it may substantially alter the intermediate representations and degrade the predictive performance of the model. We therefore recover task performance by progressively adapting the edge-side transformer blocks through knowledge distillation using the logit Kullback--Leibler divergence loss \(D_{\mathrm{KL}}\), which encourages the student edge to preserve features that reproduce the output behavior of the original model. In particular, as shown by the outer loop in Algorithm~\ref{alg:positionless_adaptation}, at the finetuning stage \(j\in\{1,\ldots,k\}\), only the parameters of the \(j\)-th student block are optimized. Previously adapted blocks remain fixed, whereas subsequent edge-side blocks retain their original pretrained parameters. The entire cloud-side network also remains fixed throughout the entire procedure, although gradients are backpropagated through it to optimize the edge-side blocks.

Notably, this progressive block-wise adaptation provides a more stable optimization of the task objective than jointly fine-tuning the entire edge-side model. In the ablation study presented in Sections \ref{sec:progressive_ablation} and \ref{sec:adversarial_ablation}, we further validate this design choice and investigate the potential benefits and limitations of incorporating a min--max optimization strategy into the formulation.

\section{Experimental Results}
\label{sec:experiments}
In this section, we first evaluate the effectiveness of the proposed attack and then assess the benefits of introducing the proposed defense against both token reduction and shuffling. Before presenting the experimental results, we describe the experimental setup.


\subsection{Experimental Settings}
\label{sec:experimental_settings}

\paragraph{Models and datasets}
We conduct all experiments on ImageNet-1K~\cite{deng2009imagenet} using ViT-B/16~\cite{Dosovitskiy21} and MAE-B/16~\cite{he2022mae}. Although the two models share the same ViT architecture, they differ in their pretraining strategies: ViT-B is trained using supervised classification, whereas MAE-B relies on self-supervised masked autoencoding and requires downstream finetuning. We therefore evaluate the pretrained ViT-B directly on ImageNet, while adapting MAE-B by fine-tuning its classification head and final Transformer block for 30 epochs on the ImageNet-1K training set, with all preceding blocks kept frozen. The latter setup also allows us to explore a split-inference scenario in which fixed client representations can support different server-side tasks. Furthermore, as shown in the following experiments, MAE-B is more vulnerable to reconstruction attacks than the supervised ViT-B, even at deeper split points, likely due to its reconstruction-oriented pretraining.

\paragraph{Token operations and metrics}
We consider token shuffling and two representative token reduction methods. Specifically, we evaluate ToMe~\cite{bolya2023tome}, which progressively merges $r$ tokens based on their similarity, and a random token dropping strategy~\cite{rao2021dynamicvit, kim2024tofu, papa2024efficientvit}, in which a fixed number $r$ of tokens is discarded after each ViT block.

To evaluate image reconstruction quality, we employ several well-established vision metrics, namely the Structural Similarity Index Measure (SSIM)~\cite{wang2004ssim}, Peak Signal-to-Noise Ratio (PSNR), and Feature Similarity Index Measure (FSIM)~\cite{zhang2011fsim}. Higher SSIM, PSNR, and FSIM values indicate better reconstruction quality and, consequently, greater information leakage and weaker privacy protection.

Finally, particularly for token reduction methods, increasing the reduction parameter may improve privacy while causing a substantial degradation in classification accuracy. To jointly evaluate privacy and utility, we introduce the \emph{Privacy--Utility Reconstruction Index} (PURI). Utility is measured as the classification accuracy relative to the unmodified baseline:
\(
U=\frac{\mathrm{Acc}}{\mathrm{Acc}_{\mathrm{base}}}.
\)
Privacy is quantified as the relative degradation in reconstruction quality compared with the baseline attack, which applies no token operations and uses a convolutional decoder at the considered split point. Accordingly, the privacy score is defined as
\begin{equation}
P=\max\left(0,1-\frac{1}{3}\left(
\frac{\mathrm{SSIM}}{\mathrm{SSIM}_{\mathrm{base}}}+
\frac{\mathrm{PSNR}}{\mathrm{PSNR}_{\mathrm{base}}}+
\frac{\mathrm{FSIM}}{\mathrm{FSIM}_{\mathrm{base}}}
\right)\right).
\end{equation}
Accordingly, $P=0$ indicates no privacy improvement over the baseline, whereas larger values indicate a stronger degradation in reconstruction quality.
Finally, PURI combines utility and privacy through a weighted harmonic mean:
\begin{equation}
\mathrm{PURI}_{\lambda}=
\frac{U \cdot P}{\lambda \cdot  P+(1-\lambda) \cdot U},
\label{eq:puri}
\end{equation}
Here, $\lambda\in[0,1]$ controls the relative importance assigned to utility and privacy. We set $\lambda=0.7$, thereby assigning greater importance to utility preservation, which reflects the practical requirement that a privacy-preserving method should maintain high classification accuracy while reducing reconstruction leakage.

\paragraph{Attack and Defense Setup}
Regarding the setup of SARA, each component is trained considering $\mathcal{D}_\mathrm{aux}$ as a randomly selected $25\%$ subset of the ImageNet-1K training set to limit computational cost. The TPP and image decoder are trained for $20$ epochs, whereas the MAE is trained for $10$ epochs. We use a learning rate of $10^{-4}$ for the TPP and MAE and $10^{-3}$ for the image decoder.


Consistent with the discussion in Section~\ref{sec:sara}, the TPP is trained on randomly shuffled smashed representations generated using different random seeds. The MAE is trained on randomly masked smashed representations to reconstruct the missing tokens, whereas the decoder is trained on the original, unshuffled and unreduced smashed representations, following the same setup as the baseline convolutional decoder.
The architectures of all SARA components are detailed in Table~\ref{tab:attack_architectures}. Their configurations were selected based on the best performance observed in preliminary experiments; for example, the TPP architecture was chosen according to the results reported in Figure \ref{fig:motivation}. Regarding the confidence threshold $\tau$ of the TPP introduced in Eq.~\ref{eq:selection_thr}, in both the attack and defense preliminary evaluation, we did not observe substantial differences in reconstruction performance when varying $\tau$. Therefore, for simplicity, all reported experiments use $\tau = 0.0$, and so every token is assigned to its most likely predicted position. 

\begin{table}[t]
\centering
\caption{Architectural setup of SARA's components.}
\label{tab:attack_architectures}
\small
\begin{tabular}{lll}
\toprule
\textbf{Component} & \textbf{Stage} & \textbf{Configuration} \\
\midrule

\multirow{3}{*}{TPP}
& Projection & Linear $768\!\rightarrow\!384$ \\
& Encoder & Transformer Encoder $\times 2$, $d=384$ \\
& Head & Linear $384\!\rightarrow\!196$ + Softmax  \\
\midrule

\multirow{3}{*}{MAE}
& Input encoding & MAE positional embedding, $d=768$ \\
& Encoder & Transformer Encoder $\times 4$, $d=768$ \\
& Head & LayerNorm + Linear $768\!\rightarrow\!768$ \\
\midrule

\multirow{3}{*}{Decoder}
& Reshape & $196\times768 \rightarrow 768\times14\times14$ \\
& Upsampling & ConvT: $768\!\rightarrow\!256\!\rightarrow\!128
\!\rightarrow\!64\!\rightarrow\!32$ \\
& Output & Conv $32\!\rightarrow\!3$ \\
\bottomrule
\end{tabular}
\end{table}

Regarding the setup used for the proposed defense, each transformer block is fine-tuned using the ImageNet-1K training set for 50 epochs with a learning rate of $10^{-4}$ and temperature $4$ in the $D_\mathrm{KL}$.
Both attack and defense performance are evaluated on the complete ImageNet-1K validation set.

Every component of the SARA pipeline is trained independently for each split point. When a defense is applied, all components of SARA are retrained with respect to the defended model.

\subsection{Attack results}
In this first part of the experiments, we evaluate SARA under token shuffling and token reduction strategies. The main objective is to assess whether our attack remains capable of reconstructing the original input despite these token operations.

\subsubsection{Shuffling}
\begin{figure}[t]
  \centering
  \includegraphics[width=0.75\columnwidth]{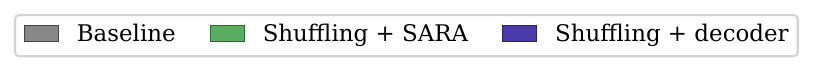}\\[2pt]
  \subcaptionbox{\small{ViT-B/16}}
    {\includegraphics[width=0.48\columnwidth]{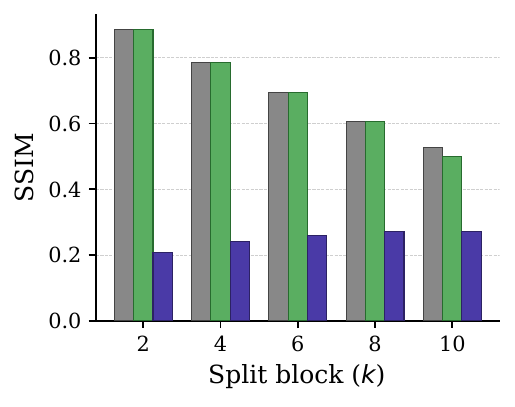}}
  \subcaptionbox{\small{MAE-B/16}}
    {\includegraphics[width=0.48\columnwidth]{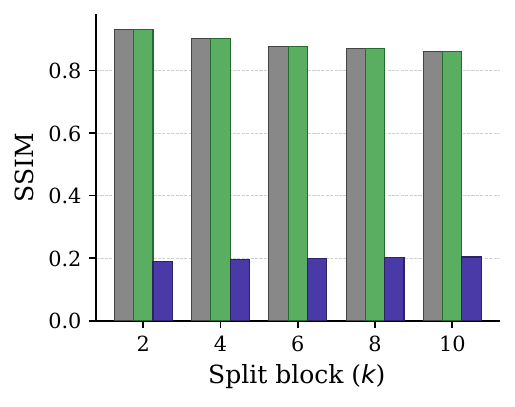}}
  \caption{\small{SSIM under token shuffling for different split blocks ($k$). \textit{Baseline} is a client with no shuffling applied (reconstruction upper bound). \textit{Shuffling + decoder} applies the convolutional decoder directly to the shuffled tokens. \textit{Shuffling + SARA} is our attack against shuffled tokens.}}
  \label{fig:ssim-splits-shuffling}
\end{figure}

As illustrated in the motivation presented in Section~\ref{sec:attack_pipeline}, inputs subjected to token shuffling cannot be reconstructed using the convolutional decoder. In contrast, as shown in Fig.~\ref{fig:ssim-splits-shuffling}, SARA can fully reconstruct the input image when token shuffling is applied to both ViT and MAE models, achieving reconstruction quality comparable to that of the baseline across multiple split points. This result is expected because token shuffling alters only the order of the tokens, while their original spatial positions can be reliably inferred even at very deep split points.

We also observe that the reconstruction quality of MAE-B/16 remains nearly constant as the split point moves to deeper blocks. This finding reinforces the observation made in the experimental setup regarding the importance of accounting for potential privacy leakage in models pretrained with reconstruction-based objectives.

\begin{takeawaybox}
\textbf{Takeaway.} Token shuffling alone does not defend against SARA, since the original token order remains recoverable even at deep ViT split points.
\end{takeawaybox}

\subsubsection{Token reductions}
\begin{figure}[t]
    \centering
    \includegraphics[width=0.75\columnwidth]{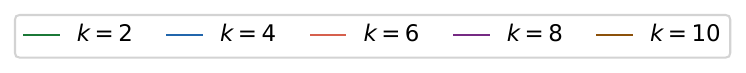}\\[2pt]
    \subcaptionbox{\small{ViT-B/16 (Dropping)}}
    {\includegraphics[width=0.48\columnwidth]{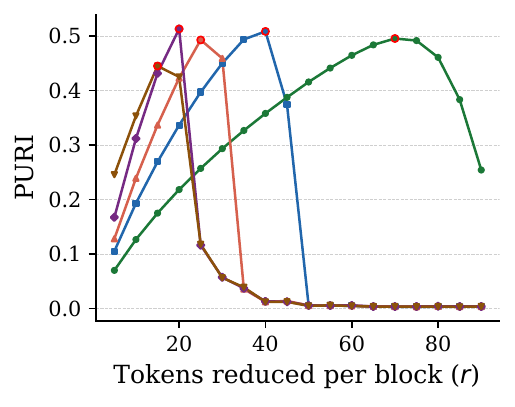}}
    \hfill
    \subcaptionbox{\small{MAE-B/16 (Dropping)}}
    {\includegraphics[width=0.48\columnwidth]{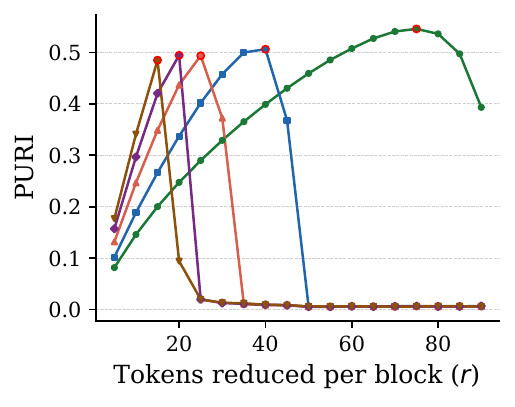}}
    \subcaptionbox{\small{ViT-B/16 (ToMe)}}
    {\includegraphics[width=0.48\columnwidth]{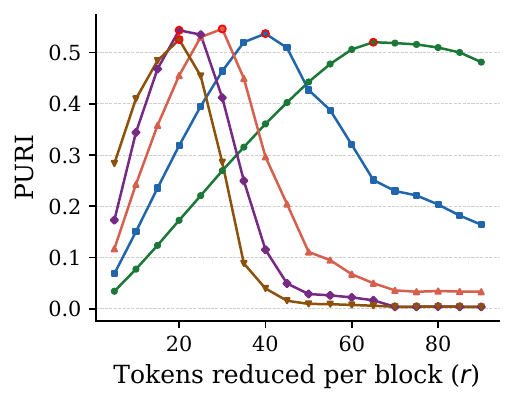}}
    \hfill
    \subcaptionbox{\small{MAE-B/16 (ToMe)}}
    {\includegraphics[width=0.48\columnwidth]{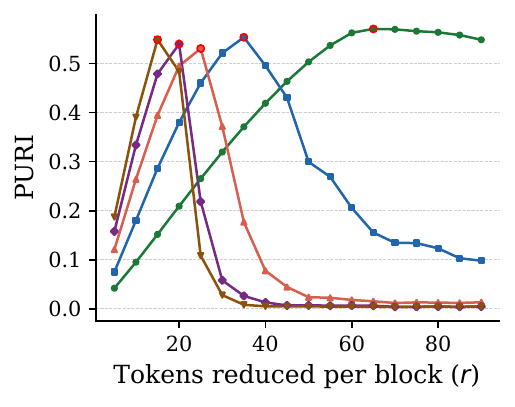}}
    \caption{\small{PURI as a function of the reduction amounts $r$ for Random Dropping (top) and ToMe (bottom), on ViT-B/16 and MAE-B/16.}}
    \label{fig:puri-r}
\end{figure}

We first evaluate how effectively SARA reconstructs inputs affected by ToMe and random token dropping under different layer-wise reduction amounts, denoted by $r$. 
Figure~\ref{fig:puri-r} reports the PURI scores, defined in Equation~\ref{eq:puri}, for reduction amounts ranging from 5 to 90. For each curve, the value of \(r\) that maximizes the PURI score is highlighted.
Figure~\ref{fig:puri-r} reveals a clear trend. At low reduction amounts, the transmitted representation remains close to the original one and also, especially for shallow split points, retains substantial information about the input. This helps preserve accuracy but provides limited privacy, resulting in a low PURI score. As the reduction amount increases, privacy improves, while accuracy gradually decreases. This degradation is generally less pronounced at shallower split points, where token reduction is applied across fewer transformer blocks.
Importantly, optimal values of $r$ at each split point identify the operating points that provide the best trade-off between downstream accuracy and privacy against reconstruction attacks. 

\begin{figure}[t]
    \centering
    \includegraphics[width=0.65\columnwidth]{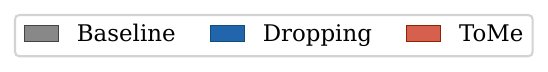}\\[2pt]
    \subcaptionbox{\small{Accuracy - ViT-B/16}}
    {\includegraphics[width=0.48\columnwidth]{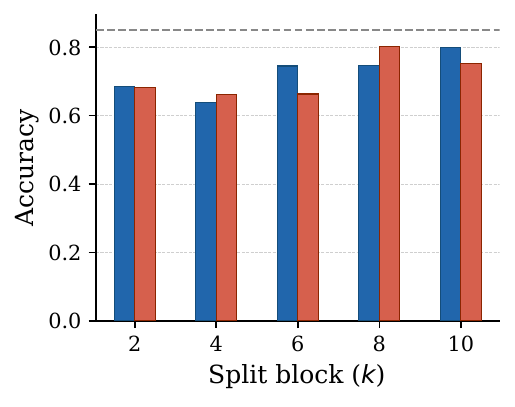}}
    \subcaptionbox{\small{Accuracy - MAE-B/16}}
    {\includegraphics[width=0.48\columnwidth]{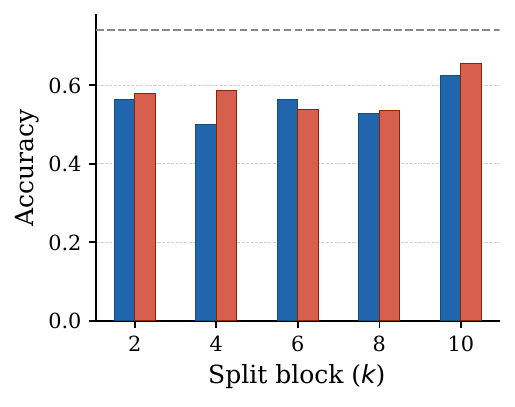}}
    \subcaptionbox{\small{SSIM - ViT-B/16}}
    {\includegraphics[width=0.48\columnwidth]{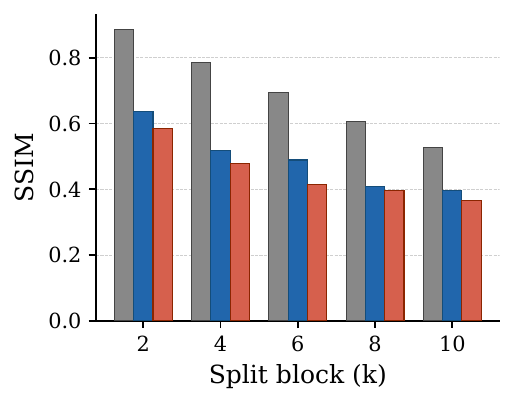}}
    \subcaptionbox{\small{SSIM - MAE-B/16}}
    {\includegraphics[width=0.48\columnwidth]{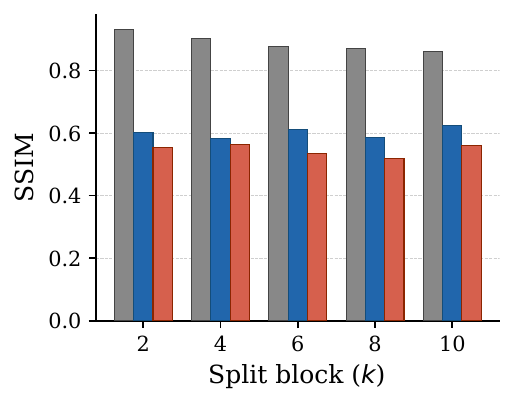}}
    \caption{\small{Classification accuracy (top row) and reconstruction quality (bottom row) at optimal $r$, on ViT-B/16 and MAE-B/16. The dashed line refers to the accuracy without reductions.}}
    \label{fig:splits-reductions}
\end{figure}

\begin{figure}[h]
\includegraphics[width=0.48\textwidth]{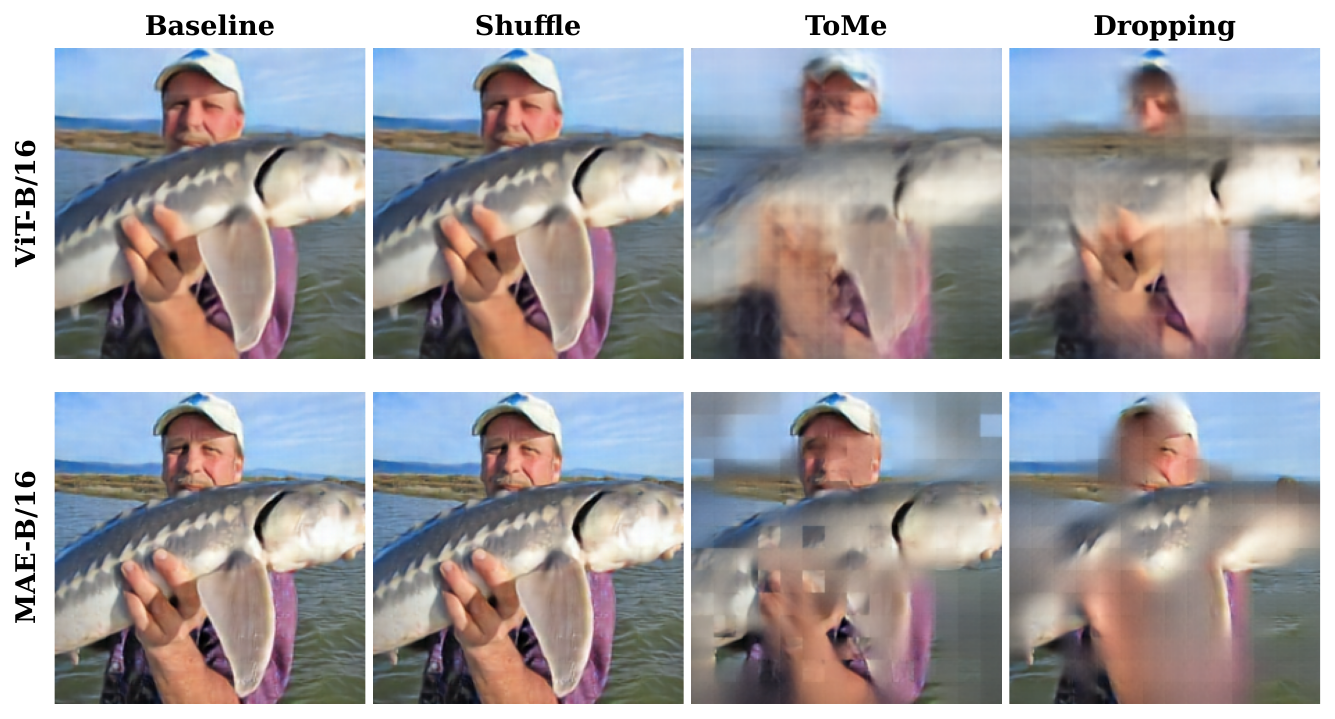}
\caption{\small{Reconstructed images from intermediate representations at split point $k=4$ for ViT-B/16 and MAE-B/16. For ToMe and Dropping, we used the optimal $r$ that maximizes PURI.}}
\label{fig:ssim-splits-reductions}
\end{figure}

We therefore use the optimal configurations as representative reduction settings to analyze reconstruction quality and downstream task performance in greater detail in Figure~\ref{fig:splits-reductions}. In this analysis, the first row shows the effect of token reduction at each split point on downstream classification accuracy. As expected, selecting the PURI-optimal setups yields a good accuracy balance, even when addressing deeper split points (where optimal $r$ is clearly lower than shallow ones, as shown in Figure~\ref{fig:puri-r}), and in all cases the accuracy stays close to the baseline value.
The second row reports instead the reconstruction quality achieved by the SARA attack, measured using SSIM. For ViT-B/16, the attack results in only moderate reconstruction degradation, particularly at the final split points of the network.
In contrast, for MAE-B/16, reconstruction quality remains remarkably stable across the network depth, with consistently high SSIM values comparable to those obtained at the second split point of ViT-B/16.
A graphical illustration across all token operations addressed is shown in Figure \ref{fig:ssim-splits-reductions}. Regarding the comparison between the two token-reduction approaches, the considered metrics do not reveal a clear distinction, likely because they are agnostic to the most semantically relevant regions of the image.

\begin{takeawaybox}
\textbf{Takeaway.} Token reduction substantially degrades SARA's reconstructions only at the later split points of ViT-B/16. In contrast, for MAE-B/16, the attack continues to produce reconstructions with an SSIM above $0.5$ across all split depths.
\end{takeawaybox}

\subsection{Defense results}
In this part of the experimental evaluation, we assess the effectiveness of the proposed defense against the SARA attack.

\subsubsection{Shuffling}
We first investigate whether combining shuffling with our defense can be effective in mitigating SARA.

\begin{figure}[t]
  \centering
  \includegraphics[width=0.75\columnwidth]{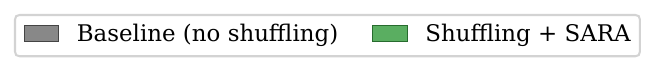}\\[2pt]
  \subcaptionbox{\small{ViT-B/16}}
    {\includegraphics[width=0.48\columnwidth]{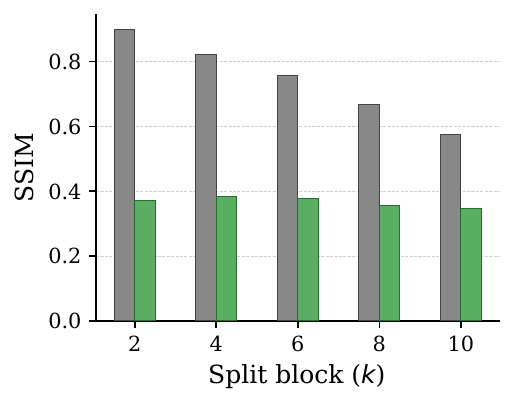}}
  \hfill
  \subcaptionbox{\small{MAE-B/16}}
    {\includegraphics[width=0.48\columnwidth]{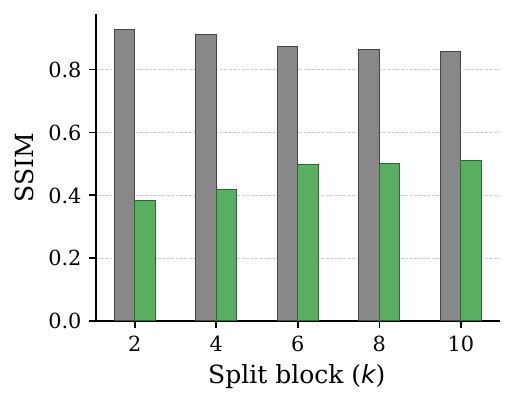}}
  \caption{\small{Reconstruction quality (SSIM) under token shuffling for different split blocks ($k$) with the proposed defense.}}
  \label{fig:def-ssim-splits-shuffling}
\end{figure}

Figure~\ref{fig:def-ssim-splits-shuffling} reports the reconstruction quality achieved by the attacker under this setting.
For ViT-B/16, the proposed defense consistently reduces the attacker's reconstruction quality across all split points, with the SSIM remaining below 0.4. This represents a substantial improvement over the baseline, indicating that attenuating the positional encoding significantly enhances the effectiveness of shuffling.
For MAE-B/16, the defense also degrades reconstruction quality, although the attack becomes more effective at deeper split points. This seemingly counterintuitive trend reflects the MAE training objective, which encourages the model to preserve spatial relationships for masked-patch reconstruction. Consequently, unlike ViT-B/16, whose deeper features become increasingly classification-specific, MAE-B/16 retains stronger spatial structure that the attacker can exploit even after positional information has been attenuated. We further study this problem and ad-hoc alternatives for the defense proposed in Section \ref{sec:positionless_adaptation}. 

\begin{figure}[t]
    \centering
    \includegraphics[width=0.75\columnwidth]{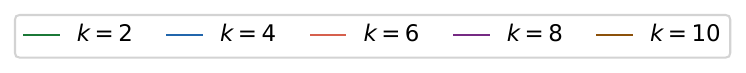}\\[2pt]
    \subcaptionbox{\small{ViT-B/16 (Dropping)}}
    {\includegraphics[width=0.48\columnwidth]{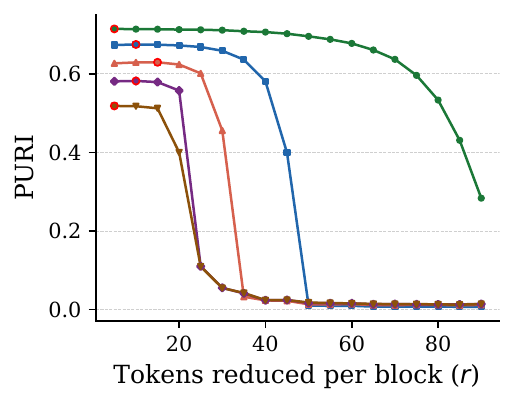}}
    \subcaptionbox{\small{MAE-B/16 (Dropping)}}
    {\includegraphics[width=0.48\columnwidth]{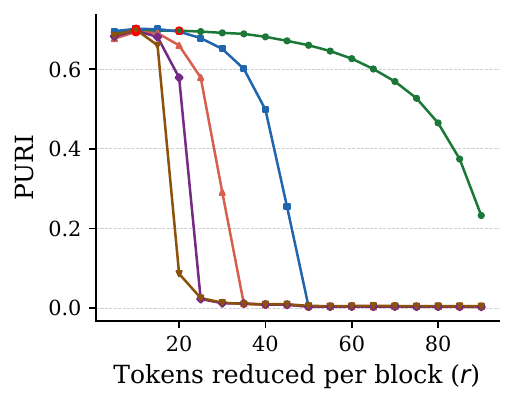}}
    \subcaptionbox{\small{ViT-B/16 (ToMe)}}
    {\includegraphics[width=0.48\columnwidth]{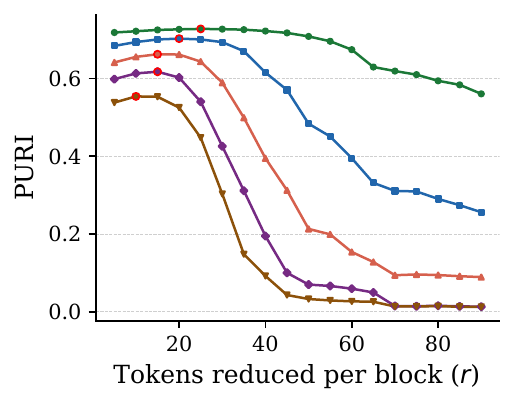}}
    \subcaptionbox{\small{MAE-B/16 (ToMe)}}
    {\includegraphics[width=0.48\columnwidth]{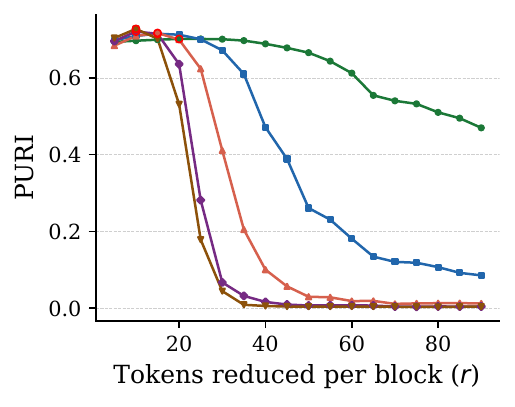}}
    \caption{\small{PURI as a function of the reduction amounts $r$ for Random Dropping (top) and ToMe (bottom), on defended models.}}
    \label{fig:puri-r-def}
\end{figure}

\subsubsection{Token reductions}
For the token-reduction analysis, we first identify the optimal \(r\) at each split point according to the PURI metric, following the procedure described in the previous experiments. The results are reported in Figure~\ref{fig:puri-r-def}, where, compared with the undefended model in Figure~\ref{fig:puri-r}, the optimal value of \(r\) is consistently lower, and the PURI curves are already close to their maximum at the smallest tested reduction amounts. This indicates that attenuating positional information substantially improves privacy against SARA even before applying aggressive token reduction. 

\begin{figure}[t]
    \centering
    \includegraphics[width=0.75\columnwidth]{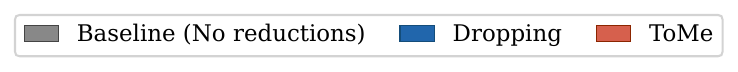}\\[2pt]
    \subcaptionbox{\small{Accuracy - ViT-B/16}}
    {\includegraphics[width=0.48\columnwidth]{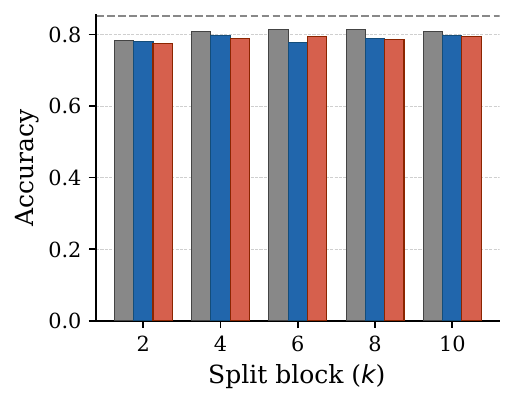}}
    \subcaptionbox{\small{Accuracy - MAE-B/16}}
    {\includegraphics[width=0.48\columnwidth]{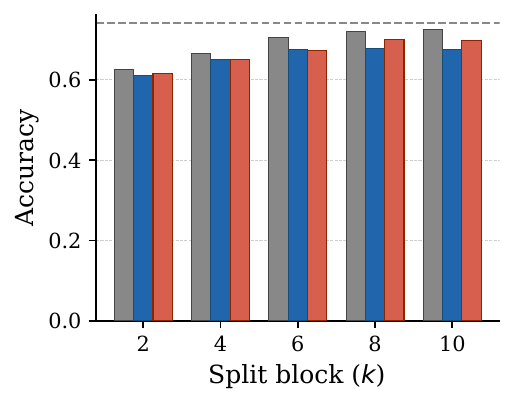}}
    \subcaptionbox{\small{SSIM - ViT-B/16}}
    {\includegraphics[width=0.48\columnwidth]{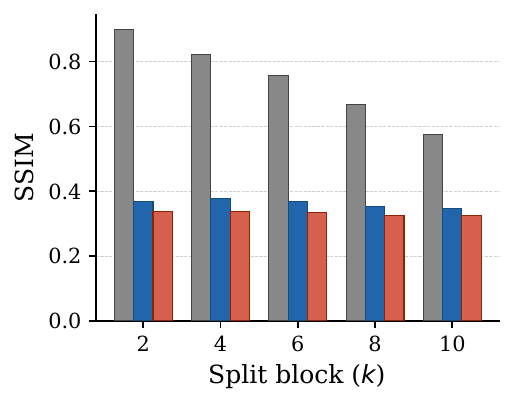}}
    \hfill
    \subcaptionbox{\small{SSIM - MAE-B/16}}
    {\includegraphics[width=0.48\columnwidth]{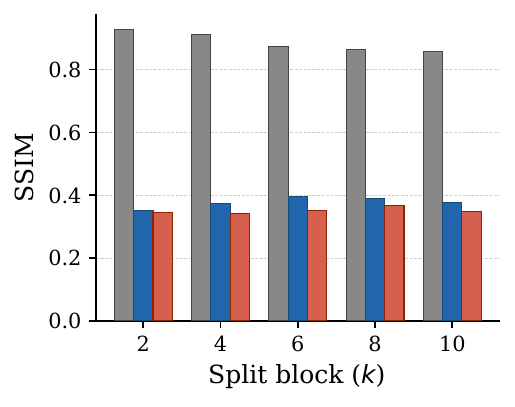}}
    \caption{\small{Accuracy (top row) and reconstruction quality (bottom row) at optimal $r$, on defended ViT-B/16 and MAE-B/16. The dashed line refers to the accuracy without defense and reductions.}}
    \label{fig:def-ssim-splits-reductions}
\end{figure}

Figure~\ref{fig:def-ssim-splits-reductions} reports the classification accuracy and reconstruction quality obtained at the optimal $r$ of each split point considered. The top panel also shows the accuracy loss introduced by the defense itself, with the dashed line indicating the performance of the original undefended model. This degradation is limited and is mainly attributable to the removal of positional embeddings from the client-side model.
Considering the accuracy drop induced by token reductions, the overall accuracy degradation is smaller than without the defense, which is mainly because the optimal reduction amounts are lower, allowing more tokens to be retained while preserving a favorable privacy--utility trade-off.

\begin{figure}[t]
\includegraphics[width=0.48\textwidth]{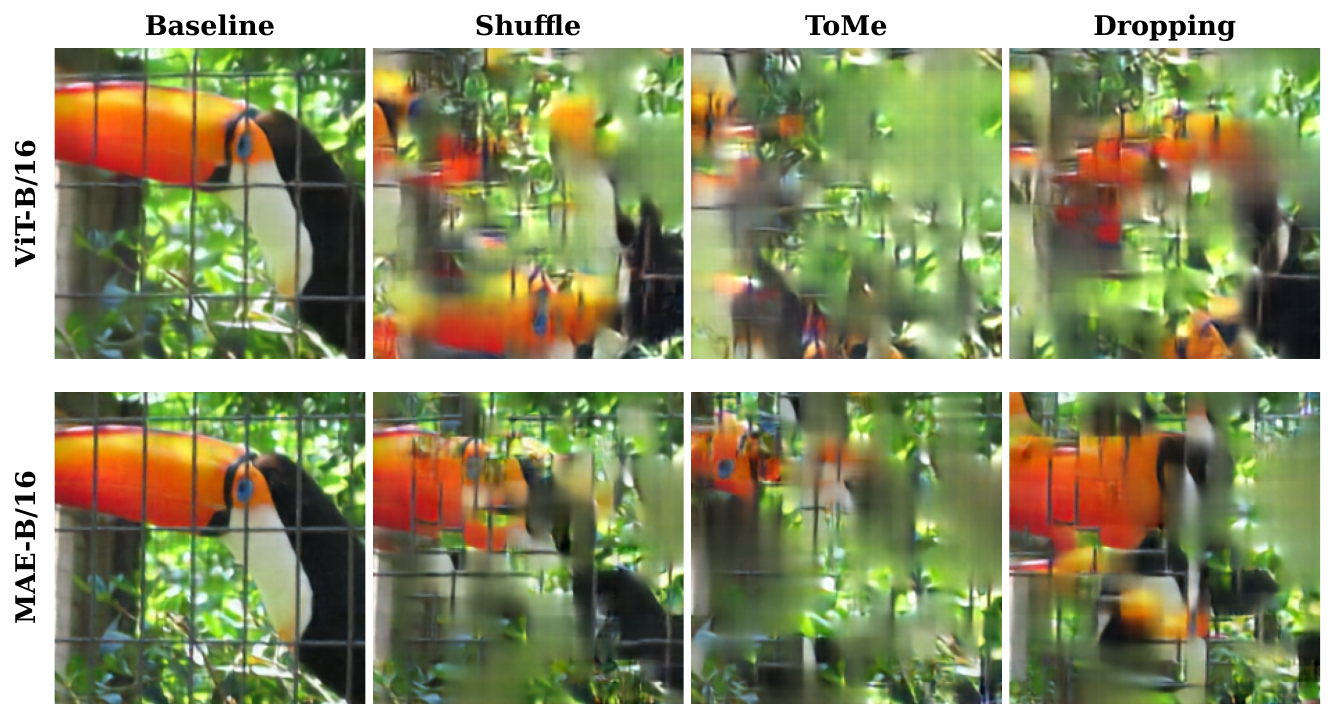}
\caption{\small{Reconstructed images from intermediate representations at split point $k=4$ for defended ViT-B/16 and MAE-B/16. Reconstructions for ToMe and Dropping are shown using the optimal $r$.}}
\label{fig:illustration_defense}
\end{figure}

As shown in the bottom panel of Figure~\ref{fig:def-ssim-splits-reductions}, the reconstruction quality remains below an SSIM of \(0.4\) across all split points for both ViT-B/16 and MAE-B/16. The increasing SSIM trend previously observed for MAE-B/16 under the shuffling-only setting remains visible but is substantially attenuated. This suggests that combining positional-information attenuation with token reduction limits the spatial information available in deeper MAE-B/16 representations, thereby reducing the effectiveness of the reconstruction attack. Illustrations of the defense benefits are shown in Figure \ref{fig:illustration_defense}.

\begin{takeawaybox}
\textbf{Takeaway.} The proposed defense introduces only a small accuracy drop relative to the original model. Under token shuffling, the defense substantially reduces the attacker's reconstruction capability across all split points, although for MAE-B/16 it slightly increases at deeper splits. When combined with token reduction, it consistently degrades reconstruction across all split points for both ViT-B/16 and MAE-B/16. 
\end{takeawaybox}

\subsubsection{Benefits of progressive finetuning}
\label{sec:progressive_ablation}
As shown in Table~\ref{tab:shuffling_puri_ablation}, the progressive finetuning strategy in  Algorithm~\ref{alg:positionless_adaptation} generally yields more stable results and a better privacy--utility trade-off than directly finetuning the entire edge-side model, even when both approaches use the same total number of training epochs. By updating one transformer block at a time, the optimization better preserves compatibility with the fixed cloud-side network and limits abrupt changes in the intermediate representations, resulting in improved performance across most split points.

\begin{table}[t]
\centering
\caption{\small{PURI scores under token shuffling. Each entry shows simple PE-removal $\rightarrow$ progressive PE-removal finetuning (Alg.\ref{alg:positionless_adaptation}).}}
\label{tab:shuffling_puri_ablation}
\setlength{\tabcolsep}{5pt}
\renewcommand{\arraystretch}{1.05}
\small

\begin{subtable}[t]{0.49\columnwidth}
\centering
\caption{ViT-B/16}
\label{tab:shuffling_puri_vit}
\begin{tabular}{c|c}
\toprule
\({k}\) & \textbf{PURI} \\
\midrule
2  & 0.705 \(\to\) \textbf{0.715} \\
4  & 0.601 \(\to\) \textbf{0.672} \\
6  & 0.573 \(\to\) \textbf{0.623} \\
8  & 0.539 \(\to\) \textbf{0.579} \\
10 & 0.488 \(\to\) \textbf{0.518} \\
\bottomrule
\end{tabular}
\end{subtable}
\hfill
\begin{subtable}[t]{0.49\columnwidth}
\centering
\caption{MAE-B/16}
\label{tab:shuffling_puri_mae}
\begin{tabular}{c|c}
\toprule
\({k}\) & \textbf{PURI} \\
\midrule
2  & \textbf{0.715} \(\to\) 0.690 \\
4  & 0.605 \(\to\) \textbf{0.685} \\
6  & 0.542 \(\to\) \textbf{0.641} \\
8  & 0.568 \(\to\) \textbf{0.638} \\
10 & 0.621 \(\to\) \textbf{0.627} \\
\bottomrule
\end{tabular}
\end{subtable}
\end{table}

\subsubsection{On Deep Layers of MAE}
\label{sec:adversarial_ablation}
The previous analysis highlighted a limitation of the proposed defense for MAE-B/16. Under token shuffling, the reconstruction quality may increase at deeper split points because MAE representations retain spatial relationships even after removing PE.
To address this limitation, we investigate an adversarial variant of the defense in Algorithm~\ref{alg:positionless_adaptation}. In addition to progressively adapting the edge-side transformer blocks without positional embeddings, we explicitly discourage the resulting smashed representations from revealing token positions. To this end, we introduce an auxiliary TPP, denoted by \(C_{\mathrm{pos}}\), which is trained to recover the original patch indices, while the student client is trained to hinder its predictions.
Let \(\bar{\theta}_{\mathrm{e}}\) and \(\theta_{\mathrm{pos}}\) denote the parameters of the student edge $\bar{f}_e$ and the auxiliary TPP, respectively. We replace the distillation-only objective in Algorithm~\ref{alg:positionless_adaptation} with the following adversarial objective, while the pipeline and progressive finetuning are the same:

\begin{equation}
    \min_{\bar{\theta}_{\mathrm{e}}}
    \max_{\theta_{\mathrm{pos}}}
    \;
    \mathbb{E}_{\mathcal{B}\subset\mathcal{D}_{\mathrm{A}}}
    \left[
        \mathcal{L}_{\mathrm{KD}}(\mathcal{B})
        -
        \lambda_{\mathrm{pos}}
        \mathcal{L}_{\mathrm{TPP}}(\mathcal{B})
    \right],
    \label{eq:adversarial_position_defense}
\end{equation}

where \(\mathcal{L}_{\mathrm{KD}}\) preserves compatibility with the fixed server-side model (same as used in line 6 of Alg.\ref{alg:positionless_adaptation}), \(\mathcal{L}_{\mathrm{TPP}}\) is the cross-entropy loss for token-position classification, and \(\lambda_{\mathrm{pos}}\) controls the trade-off between task preservation and positional-information suppression.

\begin{table}[t]
\centering
\caption{\small{Each entry reports Defense~$\rightarrow$~Defense+Adversarial. Lower SSIM and higher accuracy are better; the best values are in bold.}}
\label{tab:adversarial_comparison}
\setlength{\tabcolsep}{3pt}
\renewcommand{\arraystretch}{1.05}
\small
\begin{tabular}{c | cc | cc}
\toprule
& \multicolumn{2}{c}{\textbf{ViT-B/16}}
& \multicolumn{2}{c}{\textbf{MAE-B/16}} \\
\cmidrule(lr){2-3}\cmidrule(lr){4-5}
\(\boldsymbol{k}\)
& \textbf{SSIM} & \textbf{Acc.}
& \textbf{SSIM} & \textbf{Acc.} \\
\midrule
2
& 0.370 $\to$ \textbf{0.290}
& \textbf{78.3\%} $\to$ 71.9\%
& 0.384 $\to$ \textbf{0.355}
& 62.6\% $\to$ \textbf{63.7\%} \\

4
& 0.385 $\to$ \textbf{0.280}
& \textbf{80.9\%} $\to$ 72.2\%
& 0.419 $\to$ \textbf{0.373}
& 66.6\% $\to$ \textbf{67.4\%} \\

6
& 0.378 $\to$ \textbf{0.282}
& \textbf{81.4\%} $\to$ 70.5\%
& 0.498 $\to$ \textbf{0.390}
& \textbf{70.7\%} $\to$ 70.1\% \\

8
& 0.355 $\to$ \textbf{0.211}
& \textbf{81.3\%} $\to$ 68.4\%
& 0.501 $\to$ \textbf{0.377}
& \textbf{72.0\%} $\to$ 71.0\% \\
\bottomrule
\end{tabular}
\end{table}

\begin{figure}[t]
    \centering
    \includegraphics[width=\columnwidth]{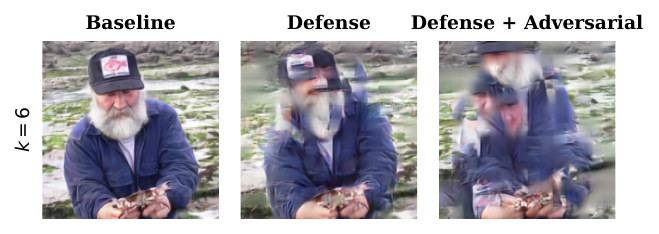}
    \vspace{-2em}
    \caption{\small{Reconstructed image in MAE-B/16 under shuffling. Comparison with baseline, defense w/o and w/ adversarial training.}}
    \label{fig:adversarial_comparison}
\end{figure}

As shown in Table~\ref{tab:adversarial_comparison}, adversarial training effectively addresses this issue by stabilizing the attacker's reconstruction capability across all ViT-MAE split points. In the proposed defense without adversarial training, the reconstruction quality under token shuffling, measured by SSIM, increases from 0.384 at split point $k=2$ to 0.501 at split point $k=8$, corresponding to an increase of $0.117$. In contrast, with adversarial training, the SSIM increases only from $0.355$ to $0.377$.
As a qualitative reference, we select split point $k=6$ to visualize the effect of adversarial training on the attacker's reconstruction quality. As shown in Fig.~\ref{fig:adversarial_comparison}, the reconstruction obtained against the adversarially trained defense is noticeably less recognizable than that obtained with the standard defense.

Despite these benefits, the adversarial formulation can introduce a non-negligible accuracy drop, particularly for ViT-B/16, due to the instability of the min--max optimization. We also omit the results for \(k=10\) from Table~\ref{tab:adversarial_comparison}, as training at this last split point exhibited stability issues. For this reason, we retain the simpler progressive finetuning strategy (Alg. \ref{alg:positionless_adaptation}) as our primary defense. A more robust integration of an adversarial regularization is left for future work.

\section{Related work}
\label{sec:related_work}
\paragraph{{Feature inversion attacks and defenses.}}
Although split inference keeps the raw input on the edge device, the intermediate representations transmitted to the cloud may still leak substantial information about the private input. This leakage can be exploited through \emph{model inversion attacks} (MIAs)~\cite{yang2020predictionpurification}, more specifically referred to as \emph{feature inversion attacks} (FIAs) in split-inference settings~\cite{he2019modelinversion, he2021attacking, salem2023sokprivacy, lei2025drag, rigaki2024surveyprivacy}. These attacks typically train a reconstruction model to map intermediate features back to the input space, often relying on convolutional decoders that exploit local spatial relationships and assume a fixed correspondence between feature locations and image regions. This assumption, however, is violated when ViT tokens are shuffled or reduced, making conventional FIAs poorly suited to evaluating manipulated token representations.

Existing defenses can broadly be grouped into cryptographic, data-modification, and learned approaches~\cite{yang2022measuring, khan2026splitattacks}. Cryptographic methods, including homomorphic encryption~\cite{dowlin2016cryptonets}, secure multiparty computation~\cite{mohassel2017secureml}, and function secret sharing~\cite{boyle2015fss}, provide strong guarantees but generally incur substantial computational and communication overhead~\cite{li2026cryptdnn, jarin2021pricure, khan2024makesplit}. Trusted Execution Environments offer a hardware-assisted alternative~\cite{sabt2015tee, wang2025splitinference}. Data-modification methods protect transmitted representations through mechanisms such as differential privacy and quantization~\cite{dwork2006differential, luo2024llrq}, often introducing a privacy--utility trade-off. Learned approaches instead optimize intermediate representations to preserve task-relevant information while suppressing sensitive content~\cite{ding2024patrol, jeong2023noisyarl, mireshghallah2021cloak}, but may require costly or unstable adversarial training. In contrast, our proposed defense operates only on the client side without introducing unstable learning trends and leaves the cloud-side model unchanged.

\paragraph{{Token-based privacy mechanisms.}}
Transformer architectures enable lightweight manipulations of intermediate representations at the token level. Token shuffling alters the order of transmitted tokens by exploiting the permutation-related properties of Transformer blocks~\cite{xu2024permutation} and has been investigated as a privacy-enhancing mechanism to hinder input reconstruction while preserving inference performance~\cite{yao2022patchshuffling}. Token reduction instead discards or merges tokens and has been primarily studied as an efficiency mechanism, with only limited analyses of its privacy implications~\cite{Erak25}. Related works have also explored the removal or anonymization of selected image regions~\cite{aslam2026pixelsprivacy} and the application of established defenses, such as differential privacy, to intermediate representations~\cite{vepakomma2018splitlearninghealthdistributed}.
Unlike cryptographic or learned defenses, token shuffling and reduction can potentially be applied without retraining or modifying the cloud-side model. However, their low overhead does not necessarily guarantee privacy.

\section{Conclusions and limitations}
\label{sec:conclusion}

In this work, we presented a comprehensive analysis of token reduction and token shuffling from a privacy perspective. We also proposed SARA, a novel attack pipeline for evaluating the privacy of these approaches. Our analysis showed that token reduction provides a certain degree of feature obfuscation. However, for token shuffling and particularly in the case of MAE-B/16, SARA was still able to produce high-quality reconstructions, highlighting the limitations of these techniques as standalone privacy-preserving mechanisms.
Motivated by these findings, we proposed a defense mechanism that attenuates positional information through a progressive fine-tuning strategy aimed at removing the positional encoding on the client side. Our experimental results demonstrate that this defense substantially improves the obfuscation of the intermediate features, reducing the effectiveness of FIAs.

Despite the promising results, we acknowledge several limitations. First, we evaluate utility only in terms of downstream classification accuracy, although the proposed fine-tuning strategy is task-agnostic. Extending the evaluation to other computer vision tasks, such as semantic segmentation, is therefore left for future work. Second, investigating whether SARA and the proposed defense can be effectively adapted to other domains, particularly language models, represents another promising direction. Finally, more sophisticated token-reordering strategies could be explored to achieve a better privacy-utility trade-off and to assess their robustness against adaptive reconstruction attacks.

\begin{acks}
This work was partially supported by project SERICS (PE00000014)
under the MUR National Recovery and Resilience Plan funded by the
European Union -- NextGenerationEU.
\end{acks}

\bibliographystyle{ACM-Reference-Format}
\bibliography{bibliography}

 \subsection*{Declaration on the Use of Generative AI}
 During the preparation of this manuscript, the authors used generative AI tools (ChatGPT-5.3 Instant and Claude Opus 4.8) exclusively to assist with grammar correction, language refinement and code generation.

\appendix
\section{Appendix}
\subsection{Optimal ratios with PURI}
Table~\ref{tab:optimal_r_comparison} reports the optimal values of $r$, within the considered range, for the two token-reduction techniques at different split points. We also report the corresponding number of transmitted tokens. For a fixed split point, increasing $r$ reduces the number of tokens transmitted from the edge to the cloud, thereby lowering communication and cloud-side computational costs, but potentially degrading task accuracy. As shown in Table~\ref{tab:optimal_r_comparison}, the PURI-optimal value of $r$ changes substantially after fine-tuning with the proposed defense. This does not imply that users should simply adopt a smaller $r$ to improve privacy. Rather, the defense shifts the privacy-utility trade-off, allowing lower reduction amounts, and thus better accuracy, while still achieving favorable privacy scores. This behavior is also evident from the comparison between the PURI curves obtained with the proposed defense and those reported in the original attack analysis. 

\begin{table}[t]
\centering
\caption{\small{Optimal reduction parameter \(r\) and corresponding number of transmitted tokens for the two architectures at \(\lambda=0.7\). Each entry reports \(r\,/\,T\), where \(T\) is the number of transmitted tokens. (a) Results without defense. (b) Results with proposed defense.}}
\label{tab:optimal_r_comparison}
\begin{subtable}{\linewidth}
\centering
\caption{Without Defense.}
\setlength{\tabcolsep}{3pt}
\renewcommand{\arraystretch}{1.05}
\footnotesize
\begin{tabular}{c | c|c | c|c}
\toprule
& \multicolumn{2}{c}{\textbf{ViT-B/16}}
& \multicolumn{2}{c}{\textbf{MAE-B/16}} \\
\cmidrule(lr){2-3}\cmidrule(lr){4-5}
\(k\) & Dropping & ToMe & Dropping & ToMe \\
\midrule
2  & 70 / 57 & 65 / 67 & 75 / 47 & 65 / 67 \\
4  & 40 / 37 & 40 / 37 & 40 / 37 & 35 / 57 \\
6  & 25 / 47 & 30 / 17 & 25 / 47 & 25 / 47 \\
8  & 20 / 37 & 20 / 37 & 20 / 37 & 20 / 37 \\
10 & 15 / 47 & 20 / 1  & 15 / 47 & 15 / 47 \\
\bottomrule
\end{tabular}
\end{subtable}
\begin{subtable}{\linewidth}
\centering
\caption{With defense.}
\setlength{\tabcolsep}{3pt}
\renewcommand{\arraystretch}{1.05}
\footnotesize
\begin{tabular}{c | c|c | c|c}
\toprule
& \multicolumn{2}{c}{\textbf{ViT-B/16}}
& \multicolumn{2}{c}{\textbf{MAE-B/16}} \\
\cmidrule(lr){2-3}\cmidrule(lr){4-5}
\(k\) & Dropping & ToMe & Dropping & ToMe \\
\midrule
2 & 5 / 187 & 25 / 147 & 20 / 157 & 20 / 157 \\
4 & 10 / 157 & 20 / 117 & 10 / 157 & 15 / 137 \\
6 & 15 / 107 & 15 / 107 & 10 / 137 & 15 / 107 \\
8 & 10 / 117 & 15 / 77 & 10 / 117 & 10 / 117 \\
10 & 5 / 147 & 10 / 97 & 10 / 97 & 10 / 97 \\
\bottomrule
\end{tabular}
\end{subtable}
\end{table}

\begin{figure}[h]
    \centering
    \begin{subfigure}[t]{0.49\columnwidth}
        \centering
        \includegraphics[width=\linewidth]{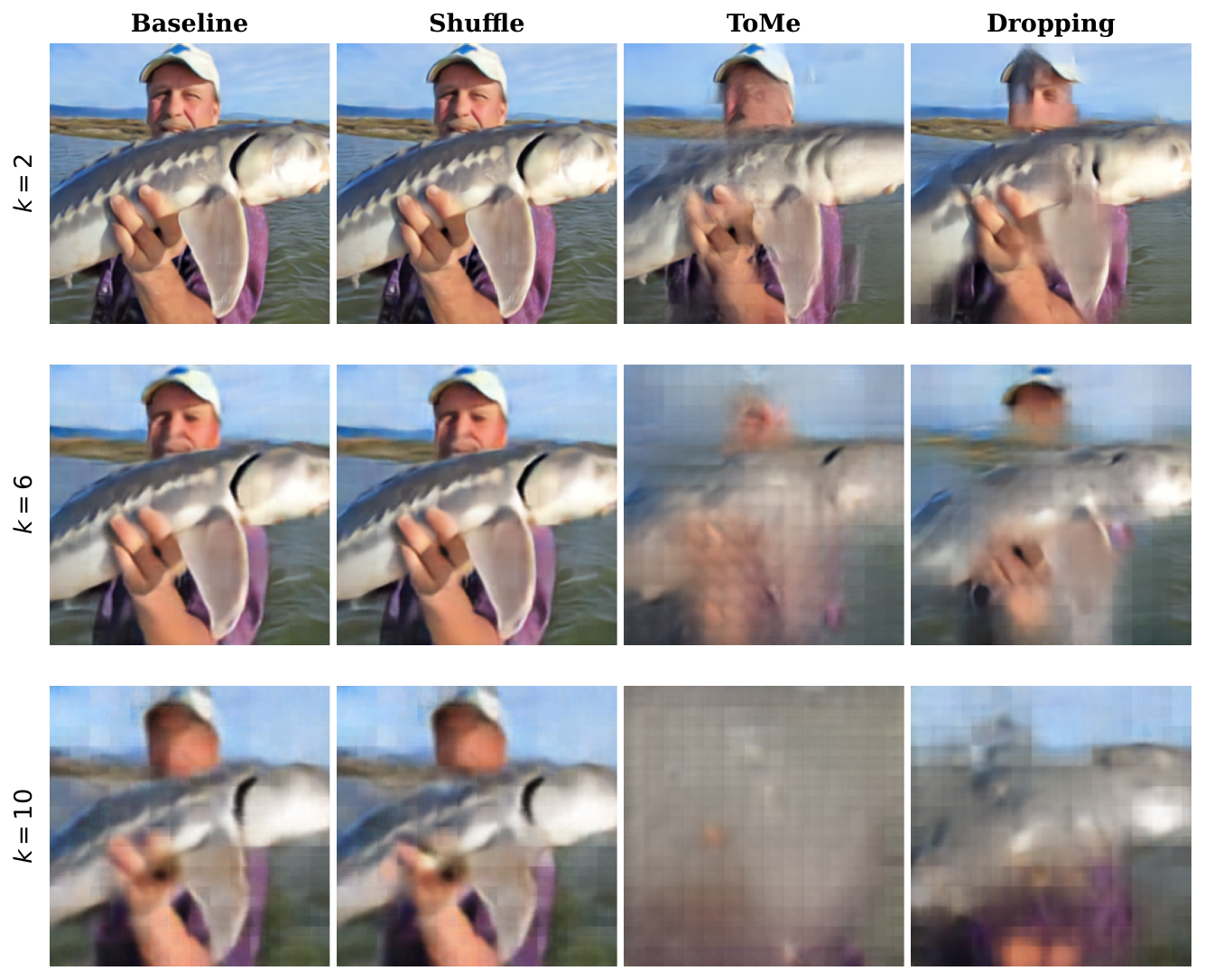}
        \caption{\textbf{ViT-B/16}}
    \end{subfigure}
    \begin{subfigure}[t]{0.49\columnwidth}
        \centering
        \includegraphics[width=\linewidth]{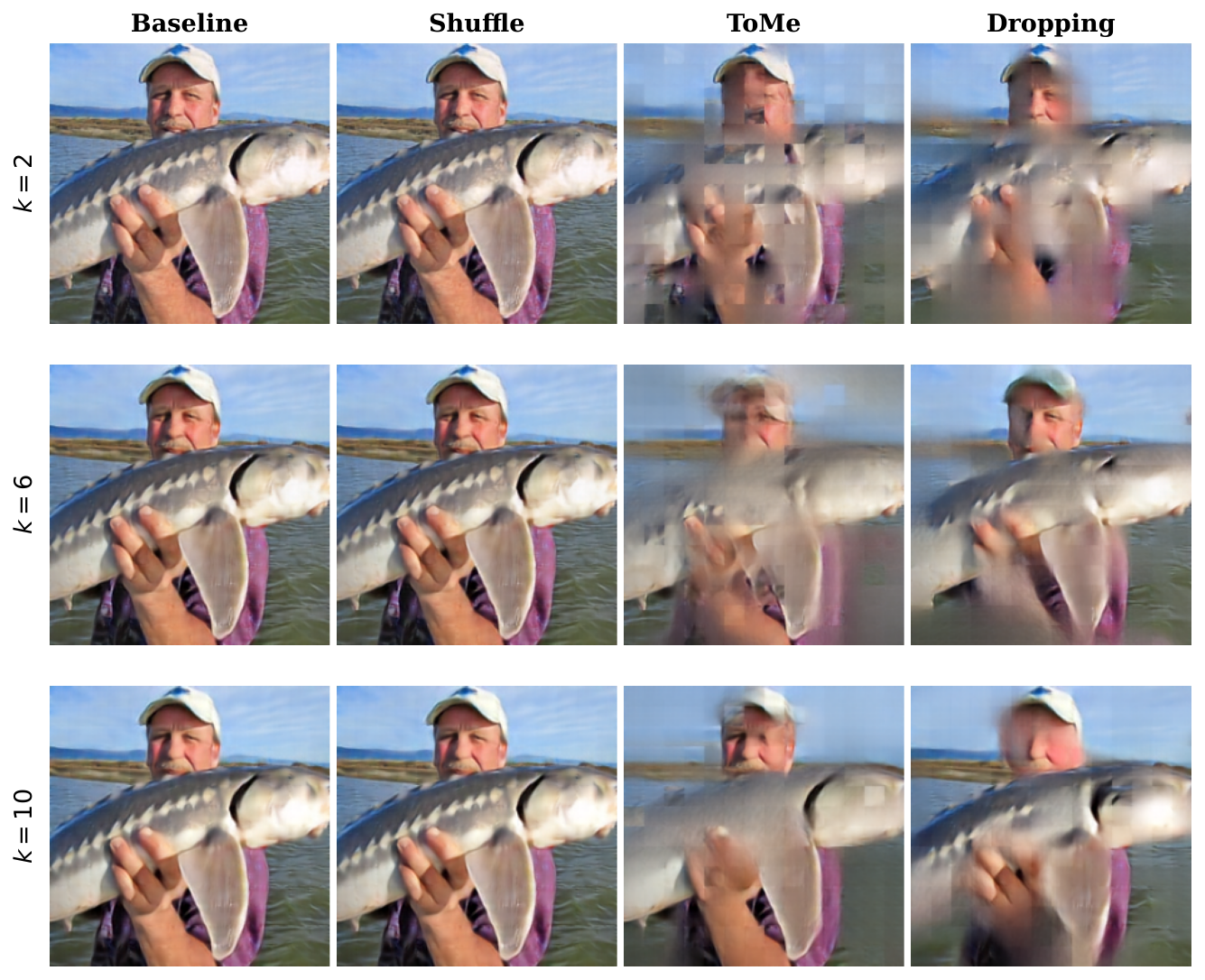}
        \caption{\textbf{MAE-B/16}}
    \end{subfigure}
    \caption{\small{Reconstructed images with SARA for baseline (no-shuffling), shuffling, and token reductions, across different split points $k$, without our defense.}}
    \label{fig:reconstructions_appendix}
\end{figure}

\begin{figure}[h]
    \centering
    \begin{subfigure}[t]{0.49\columnwidth}
        \centering
        \includegraphics[width=\linewidth]{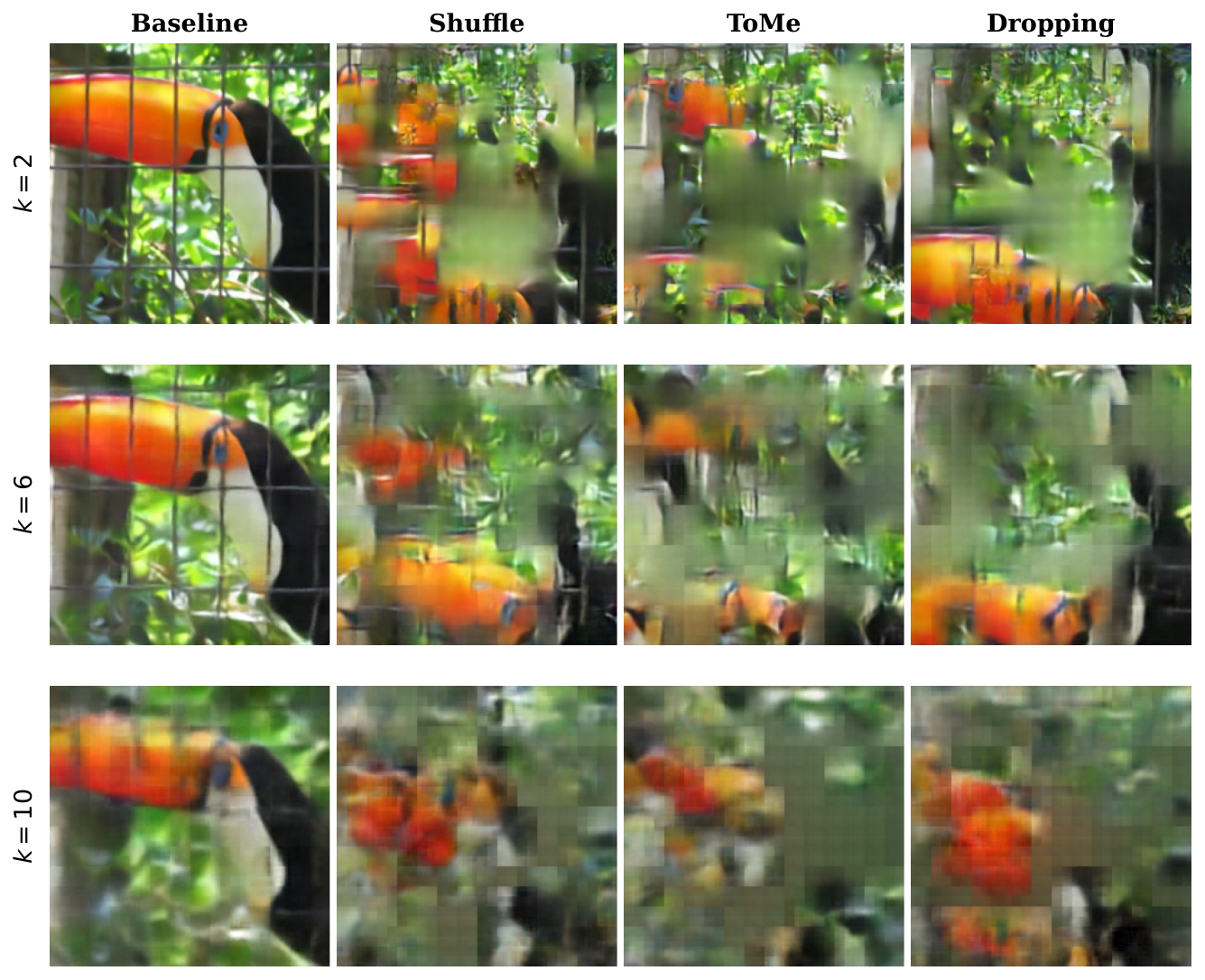}
        \caption{\textbf{ViT-B/16}}
    \end{subfigure}
    \begin{subfigure}[t]{0.49\columnwidth}
        \centering
        \includegraphics[width=\linewidth]{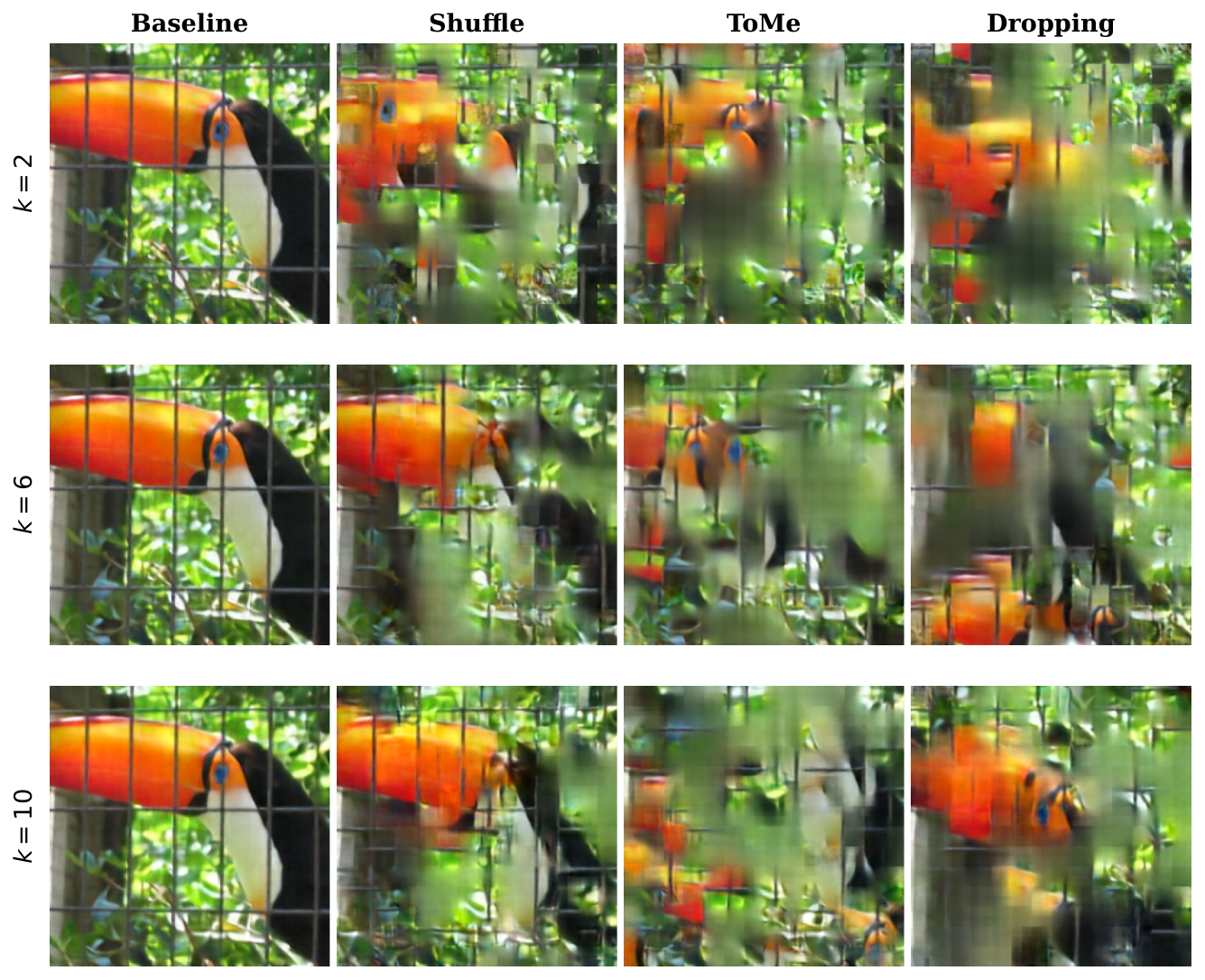}
        \caption{\textbf{MAE-B/16}}
    \end{subfigure}
    \caption{\small{Reconstructed images with SARA for baseline (no-shuffling), shuffling, and token reductions, across different split points $k$, with our defense.}}
    \label{fig:reconstructions_def_appendix}
\end{figure}

\subsection{Additional Illustrations.}
In addition to the examples presented in the main body, we provide further reconstructions in Figures \ref{fig:reconstructions_appendix} and \ref{fig:reconstructions_def_appendix} obtained at the PURI-optimal values of $r$ for both the original and defended models across different split points. Importantly, these reconstructions correspond specifically to the PURI-optimal reduction amounts.
Using a lower $r$ would allow SARA to recover higher-quality images, approaching the reconstruction quality observed under token shuffling. Conversely, a higher $r$ would make reconstruction more difficult but would also cause a greater degradation in task accuracy.

\end{document}